%% file: main.tex
\newif\ifdraft \drafttrue \draftfalse
\documentclass[a4paper, UKenglish, cleveref, autoref, thm-restate]{lipics-v2021}
\hideLIPIcs

\usepackage[utf8]{inputenc}

\usepackage{amsthm,amsmath,mathtools}
\usepackage{xargs,calc}
\usepackage{stmaryrd}
\usepackage{xcolor}
\usepackage{graphicx,tabularx}
\usepackage{pstricks}
\usepackage{xspace}
\usepackage[inline]{enumitem}
\usepackage{etoolbox,ifthen,placeins}
\newlist{enuminline}{enumerate*}{2}
\setlist[enuminline]{label={\hspace{.5ex plus .3ex minus .2ex}\rm\arabic{*})},itemjoin*={{ }and{ }}}
\newlist{subthm}{enumerate}{3}
\setlist[subthm]{label=\enumstyle{\alph{*}},ref=\thetheorem\hspace*{1pt}\enumstyle{\alph{*}}}

\newcommand{\enumstyle}[1]{{\rm(#1)}}

\usepackage{tikz}
\usetikzlibrary{arrows.meta,calc,shapes}
\input{figure/tikzstyle.tex}

\usepackage{booktabs}
\usepackage{hyperref}

\ifdraft
  \usepackage[normalem]{ulem}
  \usepackage[notref,notcite]{showkeys}
  \renewcommand*\showkeyslabelformat[1]{\scriptsize\normalfont\ttfamily #1}
\else
  \newcommand*\showkeyslabelformat[1]{}
\fi

\input{macros}

\newcommand{\OMIT}[1]{}

\newtheorem{property}[theorem]{Property}

\newtheorem*{falsestatementX}{\protect\thestatement}
\newenvironment{falsestatement}[1]{%
    \global\def\thestatement{#1}%
    \begin{falsestatementX}
}{%
    \end{falsestatementX}
    \let\thestatement\relax
}

\title{Run-Based Semantics for {RPQ}s}%

\author{Claire David}%
{Univ Gustave Eiffel, CNRS, LIGM, France}%
{claire.david@univ-eiffel.fr}%
{}%
{}%

\author{Nadime Francis}%
{Univ Gustave Eiffel, CNRS, LIGM, France}%
{nadime.francis@univ-eiffel.fr}%
{}%
{}%

\author{Victor Marsault}%
{Univ Gustave Eiffel, CNRS, LIGM, France
\and \url{https://victor.marsault.xyz}}%
{victor.marsault@univ-eiffel.fr}%
{https://orcid.org/0000-0002-2325-6004}%
{}%

\authorrunning{C. David, N. Francis, and V. Marsault}
\Copyright{Claire David, Nadime Francis, and Victor Marsault}



\ArticleNo{}

\begin{document}
\nolinenumbers

\maketitle%

\begin{abstract}
    The formalism of RPQs (regular path queries) is an important building block of most query languages for graph databases.  RPQs are generally evaluated under homomorphism semantics; in particular only the endpoints of the matched walks are returned.
    
    Practical applications often need the full matched walks to compute aggregate values.  In those cases, homomorphism semantics are not suitable since the number of matched walks can be infinite.  Hence, graph-database engines adapt the semantics of RPQs, often neglecting theoretical red flags.  For instance, the popular query language Cypher uses trail semantics, which ensures the result to be finite at the cost of making computational problems intractable.
    
    We propose a new kind of semantics for RPQs, including in particular simple-run and binding-trail semantics, as a candidate to reconcile theoretical considerations with practical aspirations. 
    Both ensure the output to be finite in a way that is compatible with homomorphism semantics: projection on endpoints coincides with homomorphism semantics.
    Hence, testing the emptiness of result is tractable, and known methods readily apply. Moreover, simple-run and binding-trail semantics support bag semantics, and enumeration of the bag of results is tractable.
\end{abstract}%

\section{Introduction}

When querying data graphs, users are not only interested in retrieving data, but also in \emph{how} these pieces of data relate to each other. This is why most languages for querying data graphs, both in theory and in practice, are \emph{navigational} languages. Informally, the querying process starts at some vertex and then \emph{walks} through the graph: it follows edges from vertex to vertex, retrieving and testing data along the way, until the walk ends in some final vertex.

In database theory, this process is usually abstracted as Regular Path Queries (RPQs~\cite{CruzMendelzonWood1987}).
An RPQ is defined by a regular expression~$R$ and is traditionally evaluated under
\emph{walk semantics} (also known as \emph{homomorphism semantics} \cite{AnglesEtAl2017}). In that case, it returns all pairs of vertices in the graph that are linked by a walk whose label conforms to $R$.
This formalism enjoys many nice properties under walk semantics and has become an important building block of most query languages over graph databases.


\begin{figure}[t]
    \newbox{\boxA}%
    \savebox{\boxA}{\input{figure/roads}}%
    \begin{minipage}[b]{\widthof{\usebox{\boxA}}}%
        \usebox{\boxA}%
    \end{minipage}%
    \hfill%
    \begin{minipage}[b]{\linewidth - \widthof{\usebox{\boxA}}}%
        \begin{align*}
            Q_1 ={}& (\rR+\rF)^* \\
            Q_2 ={}& (\rR+\rF)^*\,\rG\,(\rR+\rF)^*  
        \end{align*}%
    \end{minipage}%
    
    \vspace*{-5pt}
    
    \begin{minipage}[t]{\widthof{\usebox{\boxA}}}%
        \captionof{figure}{A graph database~$D$}%
        \label{f:runn-data}%
    \end{minipage}%
    \hfill%
    \begin{minipage}[t]{\linewidth - \widthof{\usebox{\boxA}}}%
        \captionsetup{skip=0pt}%
        \captionof{figure}{$Q_1$, a simple reachability query, and $Q_2$, reachability with a mandatory stop}%
        \label{f:runn-quer}%
    \end{minipage}%
\end{figure}

However, RPQs do not entirely meet the needs of real-life graph database systems. Indeed, limiting the output of the query to the endpoints of the walk is not enough for many real-life applications, which might also require the number of matching walks (e.g.\@ to rank answers or evaluate connectivity), or even the walks themselves (e.g.\@ for route planning)~\cite{RobinsonWebberEifrem2015}.
 Under walk semantics, the \emph{space of matches} is infinite: there are infinitely many matching walks when the graph contains cycles, which renders these questions meaningless. 
Most graph database management systems have their own way of addressing this issue, with none of them being entirely satisfactory. 
We briefly describe the most common approaches below, as well as their shortcomings,
and use the example given in Figures~\ref{f:runn-data} and~\ref{f:runn-quer}  to illustrate them.

\goodbreak

\begin{description}
\item[Topological restriction] This solution roots out unboundedness by forbidding cycles. Walks are only returned if no vertex (\emph{simple-walk semantics}) or no edge (\emph{trail semantics}) is visited twice. For instance, the language Cypher uses trail semantics \cite{FrancisEtAl2018}. 
Moreover, the new query language GQL~\cite{GQL-ISO,DeutschEtAl2022}, currently designed by ISO\footnote{ISO stands for International Standards Organisation.} to be the first standard language for property graphs, implements several topological restrictions. 

This approach aims at keeping a finite number of walks that are representative of the possibilities in the space of matches. For instance,~$Q_1$ returns $s\rightarrow t$ and $s\rightarrow c_1\rightarrow c_2\rightarrow t$ under trail semantics, two unrelated possibilities in the space of matches.
This is a crucial feature in real systems, in which pattern matching is usually just a first step before further processing.
For instance, one could evaluate the connectivity between $s$ and $t$ by counting the number of walks from~$s$ to~$t$ matching~$Q_1$.

The main weakness of this approach is that computational problems are intractable even for very simple queries. For instance, deciding whether two vertices are linked by a walk conforming to $Q_2$ is NP-complete in the size of the database under both trail~\cite{MartensNiewerthTrautner2020} and simple-walk semantics~\cite{BaganBonifatiGroz2020}. These semantics are also error-prone in that desirable results might be discarded unintentionally.
For instance,~$Q_2$ returns no walk from $s$ to~$t$ under trail semantics; and in a bigger, more realistic, graph database the walk $s\rightarrow c_1\rightarrow c_2\rightarrow c_3\rightarrow c_1\rightarrow c_2\rightarrow t$ would not be considered for further processing. 
These kinds of unwanted behaviours happen beyond theoretical settings, see \cite[p.132]{RobinsonWebberEifrem2015} for a real-life example in which the mistake occurs.

\item[Witness selection] Another approach consists in choosing a metric (length, cost, etc.), and then selecting only a few best-ranking walks in the space of matches.
For instance, the semantics of GSQL \cite{TigerGraph3.1,DeutschEtAl2019} and G-Core \cite{AnglesEtAl2018} only return the shortest walks matching the query; GQL allows returning the $k$ shortest walks \cite{DeutschEtAl2022}. 
However, the length of the path is an arbitrary metric that may not fit every application: 
here, $Q_1$ would return the ferry route $v=s\rightarrow t$ over the road route~$w=s\rightarrow c_1\rightarrow c_2\rightarrow t$ but it does not necessarily mean that~$v$ represents a faster or shorter route than~$w$ in reality.
To circumvent this issue, GQL mentions other metrics, such as $k$-cheapest, as possible extensions to investigate.
On the other hand, witness selection generally makes counting and aggregating meaningless: it counts or aggregates over something that is not representative of the space of matches.

\item[Reducing expressivity] Some systems disallow queries or operations that may lead to infinite outputs or ill-defined behaviours. Kleene stars in GQL queries are only allowed if they appear under some form of topological restriction or witness selection. In SPARQL~\cite{SPARQL1.1PP}, counting the number of walks matched by a property path is only allowed when the underlying regular language is finite. Otherwise, the returned number collapses to 0 (no walk matches the query) or 1 (at least one walk matches the query). Similarly, SPARQL equivalents of queries $Q_1$ or $Q_2$ only return the endpoints of matched walks.
Note also that switching silently from bag to set semantics depending on the query is error-prone.
\end{description}

In this article, we propose another approach called \emph{run-based}. We present two run-based semantics:
\emph{simple-run semantics}, whose input query is given as a finite automaton and provide sound theoretical foundations; and \emph{binding-trail semantics} which operate directly on a regular expression in order to be closer to practical use.
Akin to topological restriction, we aim at producing a finite output that faithfully represents the space of matches, and we do so by discarding cyclic results.
However, run-based semantics discard a result only if a cycle in the walk coincides with a cycle in the computation of the query.
For instance, binding-trail semantics filter out walks in which one edge is matched twice to the same atom of the regular expression.
Indeed, the walk $w=s\anonarrow c_1\anonarrow c_2\anonarrow c_3\anonarrow c_1\anonarrow c_2\anonarrow t$ is \textbf{not} in the output of~$Q_1$: the edge $c_1\anonarrow c_2$ is matched twice to the same $R$ atom.
On the other hand, $w$ is kept in the output
of~$Q_2$, because the two occurrences of the edge $c_1\anonarrow c_2$ are matched to two different $R$ atoms.

\medskip

The paper is organised as follows.
\rsection{prel} covers necessary preliminaries and  \rsection{run-base} gives the definition of simple-run semantics. 
%
In \rsection{comp-prob}, we revisit classical computational problems and show that simple-run semantics enjoy efficient PTIME or polynomial-delay algorithms for emptiness, tuple membership and walk enumeration. 
Counting answers remains \#P-Complete.
\rsection{quer-as-expr} defines binding-trail semantics as an adaptation of simple-run semantics to queries given as regular expressions. As a side result, we show that any regular expression (in fact, its Glushkov automaton) may encode the same behaviour and topology as any arbitrary automaton, which means that all complexity lower and upper bounds translate from one setting to the other.
Finally, we conclude this document in \rsection{brid} by a discussing possible extensions of our semantics.

\section{Preliminaries}
\lsection{prel}

\subsection{Graph databases}
\lsection{data-mode}

In this document, we model graph databases as directed, multi-labeled, multi-edge graphs, and simply refer to them as \emph{databases} for short. They are formally defined as follows:

\begin{definition}
    A \emph{database} $D$ is a tuple $(\Sigma, V, E,\src,\tgt,\lbl)$ where:%
    \begin{itemize*}[label={}]
        \item $\Sigma$ is a finite set of symbols, or \emph{labels};
        \item $V$ is a finite set of \emph{vertices};
        \item $E$ is a finite set of \emph{edges};
        \item ${\src}:E\rightarrow V$ is the \emph{source} function;
        \item ${\tgt}:E\rightarrow V$ is the \emph{target} function; and
        \item ${\lbl}:E\rightarrow 2^\Sigma$ is the \emph{labelling} function.
    \end{itemize*}
\end{definition}

Figure~\ref{f:runn-data}, page~\pageref{f:runn-data}, shows our running example for a database.

\begin{definition}
    A (directed) \emph{walk} $w$ in $D$ is a non-empty finite sequence of alternating vertices and edges of the form
    $w=(n_0,e_0,n_1,\ldots,e_{k-1},n_k)$ where~$k\geq 0$, $n_0,\ldots,n_k\in V$, $e_0,\ldots,e_{k-1}\in E$, such that:
    \begin{equation*}
        \forall i, 0\mathbin\leq i \mathbin< k, \quad \src(e_i) = n_i \quad\text{and}\quad \tgt(e_i)=n_{i+1}
    \end{equation*}
    For ease of notation, we use $\anonarrow$ to avoid naming the edge that connects two nodes when it is unique, as in $w=n_0\anonarrow n_1\anonarrow\cdots\anonarrow n_k$.

    We call~$k$ the \emph{length} of~$w$ and denote it by~$\len(w)$.
    We extend the functions $\src$, $\tgt$ and $\lbl$ to the walks in~$D$ as follows.
    For each walk~$w=(n_0,e_0,n_1,\ldots,e_{k-1},n_k)$ in~$D$,
    $\src(w) = n_0$, $\tgt(w)=n_k$, $\Endpoints(w)=\big(\src(w),\tgt(w)\big)$ and 
    \begin{equation*}\lbl(w)=\setst[\big]{u_0u_1\cdots u_{k-1}}{u_0\in\lbl(e_0),\ldots, u_{k-1}\in\lbl(e_{k-1})}.
    \end{equation*}
    Finally,~$s\rarrow{w}t$ means that~$\Endpoints(w)=(s,t)$ and, for a word $u\in\Sigma^*$, we write ${s\rarrow{u}t}$ if there exists a walk~$w$ in~$D$ such that $s\rarrow{w}t$ and $u\in\lbl(w)$.
    
    We say that two walks~$w,w'$ \emph{concatenate} if~$\tgt(w)=\src(w')$, in which case we define their \emph{concatenation} as usual, and denote it by $w\cdot w'$, or simply~$ww'$ for short.
\end{definition}

\begin{definition}
    A \emph{trail} is a walk with no repeated edge.
    A \emph{simple walk} is a walk with no repeated vertex.
    We let $\trail$ (resp.\@ $\simple$) denote the bag-to-bag function that takes as input
    a bag of walks~$B$ and returns the bag of the trails (resp.\@ simple walks) in~$B$.
\end{definition}

\subsection{Automata, expressions}
\lsection{quer}

A (nondeterministic) automaton is a 5-tuple $\Ac=\aut{\Sigma,Q,\Delta,I,F}$ where~$\Sigma$ is a finite set of symbols,~$Q$ is a finite set of \emph{states},~$I\subseteq Q$ is called the set of \emph{initial} states, $\Delta\subseteq{Q\times \Sigma\times Q}$ is the set of \emph{transitions} and~$F\subseteq Q$ is the set of \emph{final} states.
As usual, we extend~$\Delta$ into a relation over $Q\times \Sigma^* \times Q$ by adding the following: for every~$q\in Q$, $(q,\varepsilon,q)\in\Delta$; and for every~$q,q',q''\in Q$ and every~$u,v\in \Sigma^*$, if~$(q,u,q')\in\Delta$ and~$(q',u,q'')\in\Delta$ then~$(q,uv,q'')\in\Delta$.
We denote by $\lang(\Ac)$ the \emph{language} of~$\Ac$, defined as follows.
\begin{equation}
    \lang(\Ac)= \setst[\big]{u\in\Sigma^*}{\exists i\in I,~\exists f\in F,~(i,u,f)\in\Delta}
\end{equation}
A \emph{computation} in A is an alternating sequence of states and transitions that is defined similarly to walks in databases. We extend $\src$, $\lbl$, $\tgt$ and $\Endpoints$ over computations. A computation is \emph{successful} if it starts in an initial state and ends in a final~state.

A regular expression~$R$ over an alphabet~$\Sigma$ is a formula obtained inductively from the letters in~$\Sigma$, one unary function~${}^*$, and two binary functions $+$ and~$\cdot$, according to the following grammar.
\begin{equation} 
    R \mathrel{:{=}} \varepsilon \mid a \mid R^* \mid R\mathbin{\cdot}R \mid R+R   \quad\quad\text{where $a\in\Sigma$}
\end{equation}
We usually omit the $\cdot$ operator and we let~$L(R)$ denote the subset of~$\Sigma^*$ described by~$R$.

\begin{figure}[t]  
    \begin{subfigure}[b]{.2\linewidth}%
        \centering
        \input{figure/a}
        \caption{An automaton~$\Ac$}
        \label{f:ac1}
    \end{subfigure}%
    \hfill%
    \begin{subfigure}[b]{.75\linewidth}%
        \input{figure/roads-product}%
        \caption{The run database $D\times\Ac$}
        \label{f:runn-run-data}
    \end{subfigure}
    \caption{A run database constructed from $D$ (Fig.~\ref{f:runn-data}, page~\pageref{f:runn-data}) and $\Ac$ (Fig.~\ref{f:ac1}).}
    \label{f:produ-exam}
\end{figure}
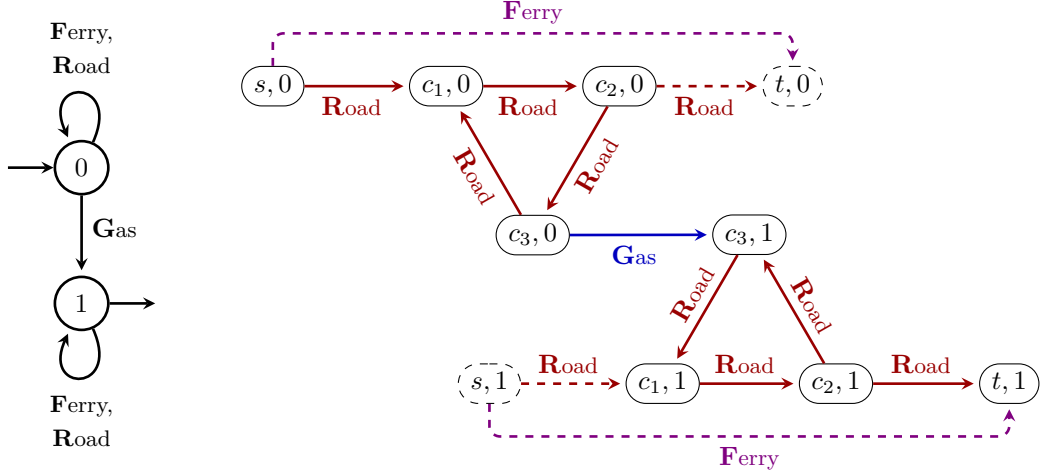
\section{Run-based query evaluation}
\lsection{run-base}

\subsection{Run-annotated database}

In this section, we fix a graph database $D = (\Sigma, V, E,\src,\tgt,\lbl)$
and an automaton $\Ac = \aut{\Sigma, Q, \Delta, I, F}$.

\begin{definition} The \emph{run-annotated database} $D\times\Ac$ (\emph{run database} for short)
is the database $D\times\Ac = (\Sigma, V', E', \src', \tgt', \lbl')$ where
    \begin{equation*}
       V' = V\times Q \quad\quad E' = \setst*{ (e,(q,a,q')) \in E\times\Delta}{a\in\lbl(e)}
    \end{equation*}
    and, for each~$e'=(e,(q,a,q'))\in E'$, 
    \begin{equation*}
        \src'(e') = \big(\src(e),q\big) \quad\quad \tgt'(e') = \big(\tgt(e),q'\big) \quad\quad \lbl'(e') = a
    \end{equation*}    
    We denote the projection from $D\times\Ac$ to~$D$ by $\projD$:
    for each $(n,q) \in V'$, $\pi_D((n,q)) = n$; for each $(e,t) \in E'$, $\pi_D((e,t)) = e$; and for each walk $w = (n_0,e_0,\ldots,n_k)$,
            $\pi_D(w) = (\pi_D(n_0),\pi_D(e_0),\ldots,\pi_D(n_k))$.
\end{definition}

The run database is essentially a product of the automaton with the database. See Figure~\ref{f:produ-exam} for an example. In the figure, unusable elements of the run database are dashed.

\begin{definition}
    A walk $w$ in $D\times\Ac$ is called \emph{a run} if $\src(w) \in V\times I$ and $\tgt(w) \in V\times F$.
    We let $\match{\Ac}{D}$ denote the bag%
    \footnote{Although the multiplicity of each element in $\match{\Ac}{D}$ is one, we would rather not use the term \emph{set} to avoid confusion when we apply bag-to-bag functions later on.}
    of all runs in $D\times\Ac$. 
\end{definition}

A simple verification yields the following property.

\begin{property}
    For every walk~$w$ in~$D$, there exists a run~$r$ in $D\times\Ac$ such that $\pi_D(r) = w$ if and only if~$\lbl(w) \cap \lang(\Ac) \neq \emptyset$.
\end{property}

\begin{remark}\lremark{automaton-changes-rundatabase}
Note that the run database~$D\times\Ac$ depends on the structure of the automaton~$\Ac$, and not only on~$\lang(\Ac)$.
Hence, when considering run databases, we usually cannot assume that~$\Ac$ is deterministic, minimal, or of any particular shape.
\end{remark}

The run database allows rephrasing the most common semantics, as in Definitions \ref{def:walk-semantics}, \ref{def:trail-semantics} and  \ref{def:simple-walk-semantics}.%

\begin{definition} \label{def:walk-semantics}
Under \emph{walk semantics}, RPQs return all walks of the input database whose label conforms to the query. 
It is defined as $\sem[W]{\Ac}(D)=\projD\compose\match{\Ac}{D}$.
\end{definition}

We will sometimes refer to the bag $\sem[W]{\Ac}(D)$ as the \emph{space of matches}, as it contains all walks that intuitively match the query. Note however that the space of matches can be infinite, and thus cannot be returned as is. 
The following two semantics circumvent this issues by restricting $\sem[W]{\Ac}(D)$ to a finite bag.

\begin{definition}\label{def:trail-semantics}
    \emph{Trail semantics} return only the trails matching the query :\\ $\sem[T]{\Ac}(D)=\trail \compose \projD \compose \match{\Ac}{D}$.
\end{definition}

\begin{definition}\label{def:simple-walk-semantics}
    \emph{Simple-walk semantics} return only the matching walks that are simple: $\sem[SW]{\Ac}(D)=\simple \compose \projD \compose \match{\Ac}{D}$.
\end{definition}

\subsection{Simple-run semantics}
\lsection{simple-run semantics}

In line with simple-walk and trail semantics, \emph{simple-run semantics}  keeps the output finite by filtering out \emph{redundant} results.
The difference between the semantics amounts to the definition of \emph{redundant}.
Classical semantics filter based on redundancy in the computed walk (repeated edge, repeated vertex), hence filtering is done \emph{after} projecting the runs to~$D$.
In the semantics we propose here, filtering is based on redundancy in the run, hence filtering is done \emph{before} projecting to~$D$.

\begin{definition}\ldefinition{simple-run semantics}
    The \emph{simple-run semantics} of an automaton $\Ac$, denoted by~$\sem[SR]{\Ac}$, is the mapping that associates, to each database $D$, the following bag of answers.
    \begin{equation}
        \sem[SR]{\Ac}(D) = \projD\compose\simple\compose\match{\Ac}{D}
    \end{equation}
\end{definition}

\begin{example}\label{ex:simple-run}
    A run in the database from Figure~\ref{f:runn-run-data} is a walk that goes from the top part to the bottom part.
    For instance, the walk $r_1={}$ $(s,0)\anonarrow(c_1,0)\anonarrow(c_2,0)\anonarrow(c_3,0)\anonarrow(c_3,1)\anonarrow(c_1,1)\anonarrow(c_2,1)\anonarrow(t,1)$ is a run, which moreover is simple.
    Hence its projection $w_1=\projD(r_1) ={}$ $s\anonarrow c_1\anonarrow c_2\anonarrow c_3\anonarrow c_3\anonarrow c_1\anonarrow c_2\anonarrow t$ belongs to $\sem[SR]{\Ac}(D)$.
    On the other hand,~$w_1$ is neither a trail nor a simple walk, hence ${w_1\notin\sem[T]{\Ac}(D)}$ and ${w_1\notin\sem[SW]{\Ac}(D)}$.
    In fact, $\sem[T]{\Ac}(D)$ and $\sem[SW]{\Ac}(D)$ contain no walk going from~$s$ to~$t$.
    
    

\end{example}


One of the main features of simple-run semantics is that it covers the space of matches, in a precise way (Lemma~\ref{l:cove-spac}).
Essentially, if a walk~$w$ matching the query is \textbf{not} returned, at least one \emph{subwalk}~$w'$ of $w$ is returned; moreover, $w'$ is obtained from $w$ by removing superfluous cycles.
Note that semantics based on topological restriction  do not enjoy the same property, as shown in Example~\ref{ex:simple-run}.

\begin{lemma}
\label{l:cove-spac}
    Let $D$ be a database, $\Ac$ be an automaton, and~$w$ be a walk in~$\match{\Ac}{D}$.
    Then, there exists a decomposition of $w$ as~$w=u_1 v_1 u_2 \cdots v_n u_{n+1}$
    such that every~$u_i$ satisfies $\src(u_i)=\tgt(u_i)$, 
    and~$v_1\cdots v_{n}\in\sem[SR]{\Ac}(D)$.
\end{lemma}
\begin{proof}
    By induction on the length of~$w$. The statement obviously holds if~$w$ is a single vertex since a walk of length 0 is always simple.
    
    Let~$w\in\match{\Ac}{D}$.
    Let~$r$ be a run in~$D\times\Ac$ such that $\projD(r)=w$.
    If $w\in\sem[SR]{\Ac}(D)$, there is nothing to prove : fix~$n=1$, $u_1,u_2$ as single vertices and~$v_1=w$.
    Otherwise, it means that~$r$ is not simple, that is there is a decomposition of $r$ as $r = r_1r_2r_3$ such that~$\tgt(r_1)=\src(r_2)=\tgt(r_2)=\src(r_3)$ and~$\len(r_2)\neq 0$.
    Hence,~$r_1r_3$ is a run in~$D\times\Ac$ and the walk~$w'=\projD(r_1r_3)$ belongs to~$\match{\Ac}{D}$.
    Then, we conclude by induction on~$w'$ and reconstruct the decomposition of~$w$.
\end{proof}

\begin{remark}\lremark{automaton-changes-simplerun}
\rremark{automaton-changes-rundatabase} shows that the run database depends on the automaton itself, and not only on the corresponding language.
This dependence carries over to simple-run semantics: $\sem[SR]{\Ac}(D)$
and $\sem[SR]{\Bc}(D)$ might be different even if $\lang(\Ac)=\lang(\Bc)$.
Choosing $\Ac$ or $\Bc$ governs which representatives of the space of matches are returned, in the sense of Lemma~\ref{l:cove-spac}.
\end{remark}

\begin{remark}
    Akin to simple-run semantics, one could define \emph{trail-run semantics} that would return the \emph{trails} of the run database. 
    While trail-run semantics would generally enjoy the same properties as simple-run semantics, the meaning of a trail in the run database is much harder to grasp. Indeed, transitions of the automaton usually have no intrinsic meaning, whereas states encode the content of the memory along the computation.
\end{remark}

\section{Computational problems}
\lsection{comp-prob}

In this section, we restate the most common computational problems related with query answering. We recall the known complexity results for the usual semantics, and give both lower and upper complexity bounds for simple-run semantics.

\subsection{Existence of a matching walk}
The problem \problemfont{Tuple Membership} consists in deciding whether there is a walk
matching the query between two given endpoints.
Under walk semantics, this problem corresponds to what is called \emph{homomorphism semantics} in most theoretical contexts, hence it is unsurprisingly tractable in that case (\rtheorem{tupl-memb-walk}).
On the other hand, \problemfont{Tuple Membership} is intractable under trail or simple-walk semantics (\rtheorem{tupl-memb-trai}).
We show in \rtheorem{tupl-memb-trac} that it is tractable under simple-run semantics.

\begin{problem}{Tuple Membership under X semantics}
    \item[Data:] A database~$D$, and a pair $(s,t)$ of vertices in~$D$.
    \item[Query:] An automaton~$\Ac$.
    \item[Question:] Does there exist a walk~$w\in\sem[X]{\Ac}(D)$ such that $\Endpoints(w)=(s,t)$?
\end{problem}

\begin{theorem}[\cite{MendelzonWood1995}]\ltheorem{tupl-memb-walk}
    \problemfont{Tuple Membership} is NL-complete under walk semantics.
\end{theorem}

\begin{theorem}[\cite{MartensNiewerthTrautner2020,BaganBonifatiGroz2020}]
    \ltheorem{tupl-memb-trai}
    \problemfont{Tuple Membership} is NP-complete under trail or simple-walk semantics. It is already NP-hard for a fixed query in both cases.
\end{theorem}

The typical query for which \problemfont{Tuple Membership} is hard under trail semantics is~$a^*ba^*$.
Hardness comes from the necessity to record which edges are matched by the left $a^*$, in order not to be matched by the right~$a^*$.
Under simple-run semantics it is not necessary to keep that record, which makes \problemfont{Tuple Membership} tractable,
as stated by \rtheorem{tupl-memb-trac}, below; it is a direct consequence of \rproposition{simplerun-contains-shortest}, which is stated afterwards and is a direct consequence of Lemma~\ref{l:cove-spac}.

\begin{theorem}\ltheorem{tupl-memb-trac}
    \problemfont{Tuple Membership} is NL-complete under simple-run semantics.
\end{theorem}

\begin{proposition}\lproposition{simplerun-contains-shortest}
    Let $D$ be a database, 
    $\Ac$ be an automaton, and $s,t$ be two vertices in~$D$.
    We let~$P_{s,t}$ denote the set $P_{s,t} = \setst[\big]{w\in\sem[W]{\Ac}(D)}{\Endpoints(w)=(s,t)}$.
    Each walk with minimal length in~$P_{s,t}$
    belongs to $\sem[SR]{\Ac}(D)$.
\end{proposition}
    

\rproposition{simplerun-contains-shortest} implies that simple-run semantics and walk semantics are equivalent for \problemfont{Tuple Membership}.
Hence known techniques for computing \problemfont{Tuple Membership} efficiently under walk semantics readily apply to simple-run semantics.
It also means that if one wants one witness for \problemfont{Tuple Membership}, one may use the shortest matching walk.





\subsection{Enumeration of matching walks}

The problem \problemfont{Query Evaluation} consists in enumerating the walks returned by the query.
It is perhaps the most important computational problem
regarding query answering since it is close to what database engines do in practice.
\problemfont{Query Evaluation} is ill-defined under walk semantics since~$\sem[W]{\Ac}(D)$ might be infinite.
Under trail or simple-walk semantics, \problemfont{Query Evaluation} is well-defined but it is intractable (\rtheorem{query-evaluation}).
By using Yen's algorithm, we show that it is tractable under simple-run semantics.

\begin{problem}{Query Evaluation under X semantics}
    \item[Data:] A database~$D$.
    \item[Query:] An automaton~$\Ac$.
    \item[Output:] All the walks $w \in \sem[X]{\Ac}(D)$.
\end{problem}

\begin{theorem}\ltheorem{query-evaluation}
    Unless P${}={}$NP, \problemfont{Query Evaluation} under trail or simple-walk semantics cannot be enumerated with polynomial-time preprocessing.
\end{theorem}

\rtheorem{query-evaluation} follows easily from \rtheorem{tupl-memb-trai}.

\begin{theorem}\ltheorem{enum-trac}
    \problemfont{Query Evaluation} under simple-run semantics can be enumerated with polynomial delay and preprocessing.
\end{theorem}
\begin{proof}[Sketch of proof]
    Computing $\sem[SR]{\Ac}(D)$ amounts to computing all simple walks from $(s,i)$ to $(t,f)$ in the run database $D\times \Ac$, for each vertices $s$ and $t$ of $D$ and each initial and final states $i$ and~$f$.
    This can be done for each $(s,i)$ and $(t,f)$ by using classical algorithms for simple-walk enumeration, such as Yen's algorithm~\cite{Yen71} (see \cite{MartensNiewerthTrautner2020} for a modern statement).
\end{proof}

\problemfont{Query Evaluation} enumerates the walks in~$\sem[X]{\Ac}(D)$, which is a bag.
Hence a walk in~$\sem[X]{\Ac}(D)$ with multiplicity~$m$ will be output~$m$ times.
We call \problemfont{Deduplicated Query Evaluation} the problem that enumerates the
\textbf{distinct} matching walks.

\begin{problem}{Deduplicated Query Evaluation under X semantics}
    \item[Data:] A database~$D$.
    \item[Query:] An automaton~$\Ac$.
    \item[Output:] All the walks $w \in \distinct(\sem[X]{\Ac}(D))$.
\end{problem}


Note that \rtheorem{query-evaluation} also holds for 
\problemfont{Deduplicated Query Evaluation} for similar reasons. We leave its complexity under simple-run semantics as an open problem.

\subsection{Counting matching walks}
\lsection{s:tupl-mult}

Counting the number of matching walks between two vertices, or \problemfont{Tuple Multiplicity}, is also important in practice. 
It corresponds to scenarios where one wants
to evaluate the connectivity between two vertices for some metrics.
\problemfont{Tuple Multiplicity} is ill-defined under walk semantics, as some tuples might have infinite multiplicity. Under the variants based on witness selection (e.g. shortest walk semantics, as explained in the introduction), the problem takes a different meaning and no longer reflects the level of connectivity between vertices.
Under trail or simple-walk semantics, this problem is known to be intractable (\rtheorem{tupl-mult-not-trac-lega}). We show that it is also the case under simple-run semantics (\rtheorem{tupl-mult-not-trac}).

\begin{problem}{Tuple Multiplicity under X semantics}
    \item[Data:] A database~$D$, and a pair $(s,t)$ of vertices in~$D$.
    \item[Query:] An automaton~$\Ac$.
    \item[Output:] 
    The total multiplicity of all walks~$w \in \sem[X]{\Ac}(D)$ s.t.\@ $\Endpoints(w)=(s,t)$.
\end{problem}

\begin{theorem}\ltheorem{tupl-mult-not-trac-lega}
    \problemfont{Tuple Multiplicity} is \#P-complete under trail and under simple-walk semantics. It is already \#P-hard in data complexity: there exists a fixed automaton $\Ac$ for which the problem is \#P-hard.
\end{theorem}

In all cases, the upper bound comes from counting the successful computations of a nondeterministic polynomial-time machine that simply guesses a trail (resp. simple walk, resp. simple run) going from $s$ to $t$ and checks that it is accepted by $\Ac$.

The hardness proof consists in an easy reduction from counting trails (or simple walks) in unlabelled graphs, two problems known to be \#P-complete \cite{Valiant1979}. 
These problems correspond to the special case where $\Ac$ is the one-state automaton equivalent to~$a^*$.
Note that for that particular~$\Ac$ it holds $\sem[SR]{\Ac}(D) = \sem[SW]{\Ac}(D)$, hence hardness of \rtheorem{tupl-mult-not-trac} follows.


\begin{theorem}\ltheorem{tupl-mult-not-trac}
    \problemfont{Tuple Multiplicity} is \#P-complete under simple-run semantics. It is already \#P-hard in data complexity: there exists a fixed automaton $\Ac$ for which the problem is \#P-hard.
\end{theorem}

    

\subsection{Walk membership}
\newcommand{\varL}{\textsf{Var}}
\newcommand{\keepL}{\textsf{Keep}}
\newcommand{\invertL}{\textsf{Invert}}
\newcommand{\evalL}{\textsf{Eval}}
\newcommand{\checkL}{\textsf{Check}}
\newcommand{\resetL}{\textsf{Reset}}

The last problem we consider here is \problemfont{Walk Membership}, which consists in deciding whether a given walk is returned.
This problem is usually considered whenever \problemfont{Tuple Membership} and \problemfont{Query Evaluation} are intractable; and as a matter of fact, it is known to be tractable for all usual semantics (\rtheorem{walk-memb-trail-trac}).
Surprisingly, it is intractable under simple-run semantics (\rtheorem{walk-memb-untr}).

\begin{problem}{Walk Membership under X semantics}
    \item[Data:] A database~$D$ and a walk~$w$.
    \item[Query:] An automaton~$\Ac$.
    \item[Question:] $w\in\sem[X]{\Ac}(D)$?
\end{problem}

\begin{theorem}\ltheorem{walk-memb-trail-trac}
    \problemfont{Walk Membership} is NL-complete under walk, trail or simple-walk semantics.
\end{theorem}
\begin{proof}[Sketch of proof] For walk semantics, the problem amounts to checking acceptance of a word in a nondeterministic finite automaton.  For trail (resp.\@ simple-walk) semantics, one has to additionally check that the input walk is a trail (resp.\@ a simple walk).  Hardness comes from an easy reduction from  ST-connectivity.
\end{proof}

\begin{figure*}%
    \centering%
    \input{figure/grid}%
    \caption{Graph encoding 3-SAT instance~$C_1\wedge C_2 \wedge C_3$ with ${C_1=\neg x_1 \vee x_3 \vee \neg x_4}$, 
    ${C_2=x_1\vee \neg x_3 \vee \neg x_4}$
    and ${C_3=x_1\vee x_2 \vee \neg x_3}$
    }
    \lfigure{reduc-graph}
\end{figure*}
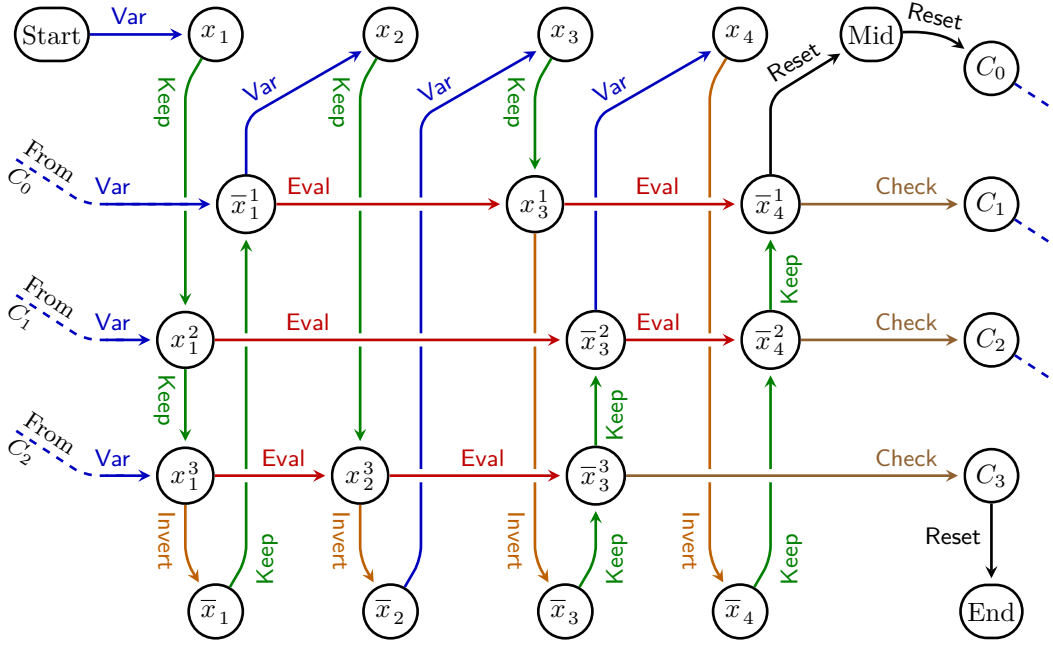

\newcommand{\walkmembuntrstatement}{\problemfont{Walk Membership} is NP-complete under simple-run semantics. It is already NP-hard in data complexity:
    there exists a fixed automaton $\Ac$ for which the problem is NP-hard.}
\begin{theorem}\ltheorem{walk-memb-untr}
    \walkmembuntrstatement
\end{theorem}
\begin{proof}[Sketch of proof]
    The hardness proof is done by a direct reduction from 3-SAT. 
    The fixed automaton $\Ac$ uses the alphabet $\{\varL, \keepL, \invertL,$ $\evalL, \checkL\}$, has three states~$\{0,1,\top\}$, $\top$ is the unique initial and final state, and its transition table is given below.
    
    \noindent{\setlength{\tabcolsep}{3pt}\def\arraystretch{0.9}%
        %
        \begin{tabular}[t]{lrcl}
        \toprule
            \keepL:   & $0$    & $\rightarrow$ & $0$ \\
              & $1$    & $\rightarrow$ & $1$ \\
              & $\top$ & $\rightarrow$ & $\top$ \\
        \midrule
            \varL:    & $\{0,1,\top\}$ & $\rightarrow$ & $\{0,1\}$ \\
        \bottomrule
        \end{tabular}
        \hspace*{0cm plus 2 fill}
        \begin{tabular}[t]{lrcc}
        \toprule
            \invertL: & $0$    & $\rightarrow$ & $1$ \\ 
              & $1$    & $\rightarrow$ & $0$ \\
        \midrule
            \resetL:  & $\{0,1,\top\}$ & $\rightarrow$ & $\top$\\
        \bottomrule
        \end{tabular}
        \hspace*{0cm plus 2 fill}
        \begin{tabular}[t]{lrcl}
        \toprule  
            \evalL:   & $1$        & $\rightarrow$ & $\{0,1\}$\\
                      & $\{0,\top\}$ & $\rightarrow$ & $\top$\\      
        \midrule
            \checkL:  & $\{0,\top\}$ & $\rightarrow$ & $\top$\\ 
        \bottomrule
        \end{tabular}
        %
    }
    
    \medskip
    
    \rfigure{reduc-graph} gives an example of how the database~$D$ is built from a specific 3-SAT instance,
    and for this example, the input walk~$w=w_V w_H$ goes through all edges of~$D$, starting with vertical edges ($w_V$) and then horizontal edges ($w_H$):
    \begin{align*}
        w_V={}&\text{Start}\rarrow{}x_1\cdots\rarrow{}\bar{x}_1\rarrow{}\cdots\rarrow{}x_2\rarrow{}\cdots \bar{x}_4 \rarrow{} \text{Mid} \\
        w_H={}&\text{Mid}\rarrow{}C_0\rarrow{}\bar{x}_1^1\cdots\rarrow{}C_1\rarrow{}\bar{x}^2_1\rarrow{}\cdots\rarrow{}C_3\rarrow{}\text{End}
    \end{align*}
    Each valuation of the variables corresponds in a one-to-one manner to one run in~$D\times\Ac$ for the vertical part.
    For instance, valuation~$x_1\mapsto 1$, $x_2\mapsto 1$, $x_3\mapsto 0$, $x_4\mapsto 0$ corresponds to the run~$r_V$:
    \begin{equation*}\arraycolsep=3pt
        \begin{array}{@{}lccccccccccccccccccccccccccc@{}}
        \text{In }D:   & \text{Start} & x_1 & x_1^2 & x_1^3 & \bar{x}_1 & \bar{x}^1_1 & x_2 & {\cdots} & x_3 & \cdots & x_4 & \cdots & \text{Mid}
        \\
        \text{In }\Ac: & \top & 1 & 1 & 1 & 0 &0 & 1 & {\cdots} & 0 & \cdots & 0 & \cdots & \top
        \end{array}
    \end{equation*}
    The horizontal walk~$w_H$ has three parts ($C_0\rarrow{}C_1$, $C_1\rarrow{}C_2$ and $C_2\rarrow{}C_3$);
    each~$C_{i-1}\rarrow{}C_i$ checks whether the valuation makes the clause~$C_i$ true.
    Let us take~$C_1$ for instance.
    There is exactly one run~$r_1$ for the part~$C_0\rarrow{} C_1$ in order for
    $r_Vr_1$ to be simple, given below.
    \begin{equation*}\arraycolsep=3pt
        \begin{array}{@{}lcccccccccccc@{}}
        \text{In }D:   & C_0 & \bar{x}_1^1 & x_3^1 & \bar{x}_4^1 & C_1    
        \\
        \text{In }\Ac: & \top & 1 & 1 & 0 & \top
        \end{array}
    \end{equation*}
    Indeed, the state reached at~$\bar{x}_1^1$ is necessarily~$1$, otherwise the full run would not be simple: the vertex~$(\bar{x}_1^1,0)$ of the run database was already visited in the vertical part.  One may see that the state reached at vertex~$C_1$ is~$\top$, which means that the valuation satisfies~$C_1$.
    If the valuation did not make~$C_1$ true, there would be no run~$r_1$ such that $r_Vr_1$ is simple.

    See Appendix~\ref{a:walk-memb} for a complete proof.
\end{proof}

    

\section{Query given as a regular expression}
\lsection{quer-as-expr}

In \rsection{simple-run semantics}, we defined simple-run semantics for queries given as automata. In this section, we show how our setting can be adapted to handle queries given as regular expressions, which is more in line with real-life query languages.
\rsection{e-to-a} discusses translating expressions to automata and explains why the resulting behaviour might not be ideal. On the contrary, \rsection{binding-trail} adapts simple-run semantics in order to work directly on a regular expression.

\subsection{Expression to automaton}
\lsection{e-to-a}

There are many known techniques for producing automata that are equivalent to a given regular expression (see for instance \cite{Sakarovitch2021}). 
The main advantage of this approach is that all algorithms from \rsection{comp-prob} immediately apply.
However, the crux of the matter lies in \rremark{automaton-changes-rundatabase}: semantics do not only depend on the language accepted by the automaton, but on the automaton itself. 
Thus, \emph{how} we choose to translate the expression into an equivalent automaton matters.
In fact, it seems that each translation algorithm features undesirable quirks.
We give two compelling examples, and leave a complete account of the many translation algorithms for future work.

First, one could translate the expression into the equivalent minimal DFA, but this choice makes the semantics non-compositional. 
Consider an expression~$R=R_1+R_2$, one would expect~$R$ to return more results than~$R_1$ or~$R_2$; but it is not always the case.
For instance, if~$R_1=b^*(ab^*ab^*)^*$ and~$R_2=(a+b)(a+b)^*$,
then~$R$ is equivalent to~$(a+b)^*$.
The minimal DFA~$\Ac_1$ associated with~$R_1$ has two states, while the minimal DFA~$\Ac$ associated with~$R$ has only one state.
Hence one can easily find a database $D$ such that $\sem[SR]{\Ac}(D)\not\supseteq\sem[SR]{\Ac_1}(D)$.
Similar quirks can be found for concatenation and star.
This example illustrates that the translated automaton must not only represent the \emph{regular language} but also the \emph{regular expression as written}.


%

Second, one could use Glushkov construction (recalled in Definition~\ref{def:glushkov}, below), which famously produces an automaton that stays close to the regular expression.
However, this choice introduces a left-to-right bias.
Typically, $\Ac=\gl(a^*b^*)$ is not the mirrored automaton of $\gl(b^*a^*)$, and if one considers the following database~$D$,
the walk $\mathtt{S}\anonarrow\mathtt{S}\anonarrow\mathtt{T}$ belongs to $\sem[SR]{\Ac}(D)$ but not the walk $\mathtt{S}\anonarrow\mathtt{T}\anonarrow\mathtt{T}$.

\hspace*{0cm plus 1fil}\input{figure/asapbbs}

\begin{definition}\label{def:glushkov}
    Let $R$ be a regular expression over $\Sigma$. A \emph{linearisation} of $R$ is a copy $R'$ of $R$ in which each atom in $\Sigma$ is replaced by a different symbol in a new alphabet $\Gamma$, called the \emph{positions} of $R$. Given $\alpha \in \Gamma$, we denote as $\overline \alpha$ the label of its antecedent in~$R$.
    
    The \emph{Glushkov's automaton} of $R$, $\gl(R)$, is the trim of the automaton $\langle \Sigma, \set{i} \uplus \Gamma, \Delta, \set{i}, F \rangle$ defined as follows:
    \begin{align*}
        \Delta ={}& \setst{(\alpha,\overline \beta,\beta)}{\exists u,v\in\Gamma^*, u\alpha \beta v \in L(R')} \cup \setst{(i,\overline \alpha,\alpha)}{\exists u\in\Gamma^*, \alpha u \in L(R')}
        \displaybreak[0]
        \\
        F ={}& \setst{\alpha}{\exists u\in\Gamma^*, u \alpha \in L(R')} \cup \set{i} \textrm{ if }\varepsilon \in L(R')
    \end{align*}
\end{definition}
%

\subsection{Binding-trail semantics}
\lsection{binding-trail}

This section defines \emph{binding-trail semantics}, a counterpart to simple-run semantics that operates directly on a given regular expression $R$, without translating $R$ into an automaton.

\begin{definition}\ldefinition{bind-trai}
    Let $D=(\Sigma, V, E,\src,\tgt,\lbl)$ be a database, and~$R$ a regular expression. Let $R'$ be a linearisation of $R$ and $\Gamma$ be the corresponding positions of $R$.
    A \emph{binding trail} of $D$ matching $R$ is a sequence $(e_1,\alpha_1)\ldots(e_n,\alpha_n)$ of pairs in $E\times\Gamma$ such that:
    \begin{itemize}
        \item $e_1\ldots{}e_n$ describes a walk of $D$ and $\overline{{\alpha_1}\cdots{\alpha_n}} \in \lbl(e_1\ldots e_n)$;
        \item $\alpha_1\cdots \alpha_n$ belongs to $L(R')$;
        \item All $(e_i,\alpha_i)$ are pairwise distinct.
    \end{itemize}
    
    \noindent Binding-trail semantics are then defined by:
    \begin{equation*}
        \sem[BT]{R}(D) = \projD\big(\setst{t}{ t\text{ is a binding trail of }D\text{ matching R}}\big)
    \end{equation*}
\end{definition}

In other words, given a walk~$w$ in~$D$,
$w$ conforms to binding-trail semantics if~$w$ matches $R$ in such a way that the same edge of~$w$ cannot be used twice at the same position in $R$.
The following lemma shows how binding-trail semantics relate to the run database.

\begin{lemma}\llemma{cara-bind-trai}
For every regular expression~$R$ and database~$D$,
    \begin{equation*}
    \sem[BT]{R}(D) = \pi_D \compose \bindingtrail \compose \match{\Ac}{D}~,
    \end{equation*}
    where~$\Ac=\gl(R)$ and~$\bindingtrail$ is the run-bag filter that keeps only the runs 
    \begin{equation*}
        (n_0,q_0)(e_0,\delta_0)(n_1,q_1)\cdots(e_{k-1},\delta_{k-1})(n_k,q_k)
    \end{equation*} such that all $(e_i,q_{i+1})$'s, $0\leq i < k$, are pairwise distinct.
\end{lemma}

As is the case for simple-run semantics, binding-trail semantics also coincide with walk semantics for \problemfont{Tuple Membership}, and produces the same shortest witnesses. Indeed, the proof of \rproposition{simplerun-contains-shortest} can easily be adapted to prove the following:

\begin{proposition}\lproposition{bindingtrail-contains-shortest}
    Let $D$ be a database and 
    $R$ be an expression. Let $\Ac$ be any automaton such that $L(R) = L(\Ac)$.
    Let~$s,t$ be two vertices in~$D$,
    and we denote by~$P$ the set of walks in $\sem[W]{\Ac}(D)$ that go from~$s$ to~$t$.
    Let~$w$ be a walk in~$P$ with minimal length.
    Then,~$w$ belongs to~$\sem[BT]{R}(D)$.
\end{proposition}

\rlemma{cara-bind-trai} hints at the fact that the upper bounds of \rsection{comp-prob} for simple-run semantics immediately apply to binding-trail semantics, due to standard graph reduction techniques translating vertex-disjoint walks to edge-disjoint walks and back. The same does not necessarily hold true for lower bounds. Glushkov automata have additional properties: only one initial state, all incoming edges to a given state have the same label, and so on (see \cite{CaronZiadi2000}). 
We show that these properties cannot be used to design more efficient algorithms.

\newcommand{\bindingtrailsimplerunequivalencesttmt}{\problemfont{Tuple Membership}, \problemfont{Tuple Multiplicity}, \problemfont{Query Evaluation} and  \problemfont{Walk Membership} are computationally equivalent under binding-trail semantics and under simple-run semantics.}
\begin{proposition}\lproposition{bindingtrail-simplerun-equivalence}
    \bindingtrailsimplerunequivalencesttmt
\end{proposition}

\rproposition{bindingtrail-simplerun-equivalence} is actually a consequence of a much deeper result stating essentially that, given any automaton $\Ac$, there exists a regular expression $R$ that encodes the topology of $\Ac$ in the sense that there is a strong connection between the computations of $\Ac$ and those of $\gl(R)$. Hence, any problem that takes automata as input will likely have hard instances that are of the form $\gl(R)$ for some expression $R$.

Due to space constraints, we only sketch the main idea of the encoding. A precise statement of the result, along with necessary definitions and proofs are given in full details in Appendix \ref{a.topo_coding}.

\begin{proof}[Sketch of proof]
Let~$\Ac\stdaut{}$.
    Let~$m=\card(\Delta)$ and $G$ denote any bijection $G:\Delta\rightarrow\set{1,\ldots,m}$.
    Let $H$ be the only bijection $H:\Delta\rightarrow\set{1,\ldots,m}$ that meets:
    $\forall e\in\Delta,~G(e)+H(e)=m+1$.
    Finally, let $\sigma$ be a fresh symbol that is not in $\Sigma$.
    We define the expression~$R$ over the alphabet $\Sigma\uplus\set{\sigma}$ as follows:
    \begin{equation*}\label{eq:R_A}
        \left(
        \sum_{q\in Q}\Bigg[
            \Big(
                \overbrace{\rule{0pt}{2ex}\varepsilon~~+}^{\text{~if $q\in I$}}
                \sum_{\substack{s\in Q,~a\in\varSigma\\ e=(s,a,q)\in \Delta}} a^{G(e)}
            \Big)
            ~\sigma~
            \Big(
                \overbrace{\rule{0pt}{2ex}\varepsilon~~+}^{\text{~if $q\in F$}}
                \sum_{\substack{a\in\varSigma,~t\in Q\\ e=(q,a,t)\in \Delta}} a^{H(e)}        
            \Big)
        \Bigg]
        \right)^*
    \end{equation*}
    
    Note that there are exactly $\card(Q)$ occurrences of the letter~$\sigma$ in~$R$.
    We associate each state~$s\in Q$ with the occurrence of~$\sigma$ appearing in the term of the external sum when $q = s$. 
    Similarly, each transition~$e = (s,a,t)$ with label $a$ in~$\Ac$ is encoded by the word $\sigma a^{m+1}\sigma$ that will be matched by the concatenation of  the $\sigma$ corresponding to $s$, the subexpression~$a^{H(e)}$ on its right, the $a^{G(e)}$ on the left of~$\sigma$ corresponding to $t$ followed by this $\sigma$.

    We conclude by showing that a successful computation in~$\Ac$ over a word $a_0\cdots a_n\in \Sigma^*$ is encoded by matching the word 
    $\sigma{a_0}^{m+1}\sigma\cdots{a_n}^{m+1}\sigma$ in $R$. Moreover, this encoding preserves relevant topological properties. For instance, a computation of $\Ac$ reuses a state if and only if its encoding reuses a position labelled by $\sigma$.
\end{proof}

\begin{remark}
    Similarly to simple-run semantics (see \rremark{automaton-changes-simplerun}), binding-trail semantics depend on the given regular expression and not only on the corresponding language. This allows for a finer control on which repetitions are permitted by the query. For instance, the three following expressions have different meanings under binding-trail semantics:
    \begin{itemize}
        \item $\sem[BT]{a^*}$ returns all trails labelled by $a$.
        \item $\sem[BT]{a^*\cdot a^*}$ returns all walks that are the concatenation of two trails.
        \item $\sem[BT]{(a + a)^*}$ returns all walks where edges are repeated at most twice.
    \end{itemize}
\end{remark}

\section{Perspectives}
\lsection{brid}

Run-based semantics are an attempt at addressing real-life concerns while maintaining good theoretical foundations. As such, our medium-term goal is to make sure that our work is indeed applicable to query languages used in practice. GQL offers a very plausible opportunity for integration, as it is in active development and already supports several semantics. This section aims at closing some of the gaps between theory and practice by discussing how run-based semantics adapt to commonly seen extensions or limitations.

\subsection{Syntax restrictions used in practice}

Real query languages, such as GQL or Cypher, impose syntax
restrictions on the regular expression given as input. We discuss
here whether the lower bound complexity results change
by imposing those restrictions. Note that no reasonable syntax
restriction can change the complexity of \problemfont{Tuple Multiplicity} since
the lower bound already holds for the fixed expression $a^*$.

We consider three syntax restrictions and their respective impact on
the complexity of \problemfont{Walk Membership}.
A regular expression has \emph{star-height 1} if it has no nested Kleene stars.
A regular expression is said to have \emph{no union under star} (resp. \emph{no concatenation under star}) if no union (resp. no concatenation) operator occurs in any subexpression nested under a Kleene star.

These restrictions match syntax rules commonly seen in practice: GQL only allows expressions with star-height 1 and Cypher queries cannot express concatenations under star. Moreover, users tend not to use the full expressive power at their disposal. In \cite{BonifatiMartensTimm2020}, the authors make an analytical study of over 240,000 SPARQL queries: in the collected data set, every single RPQ has star-height 1, all queries but one have no concatenation under star, and only
about 40\% of queries use a union under a star.

\medskip

The expression shown in \rsection{e-to-a} to simulate the computations of any arbitrary automaton has no nested stars. Hence all complexity lower bounds hold even when expressions are restricted to star-height~1.
That expression has unions under star, but one can design another expression with similar properties that does not. Indeed, let~$\Ac\stdaut{}$ be an automaton; for simplicity we assume that~$\Sigma=\{0,1,\ldots,k-1\}$ and that~$Q=\{0,1,\ldots,n-1\}$. Consider the following expression over the alphabet~$\set{a,b,c,\sigma}$.
\begin{equation}\lequation{glus-redu-no-unio-unde-star}
    \underset{i\in I}{\Pi}  c^{i+1}
    \left(
    \underset{i\in Q}{\Pi}\left(c^{n-i}\sigma a^{i+1}\right)^*
    \cdot
    \underset{(i,x,j)\in\Delta}{\Pi}
    \left(a^{n-i}b^{x}c^{j+1}\right)^*
    \right)^*
    \underset{i\in F}{\Pi}  a^{n-i}
\end{equation}
As in \rsection{binding-trail}, one can show that this new expression encodes the behaviour of $\Ac$. The key arguments are as follows: each letter~$x\in \Sigma$ is encoded by~$\lambda(x)=a^{n+1} b^x c^{n+1}\sigma$; each word $x_0\cdots x_n$ is encoded by~$c^{n+1}\lambda(x_1)\cdots\lambda(x_n)a^{n+1}$; the states of $\Ac$ are simulated by the positions with label $\sigma$.

\begin{remark}
The internal Kleene stars in \requation{glus-redu-no-unio-unde-star} are used at most once, so these stars might be replaced by an \emph{optional} operator, sometimes denoted by a ``\texttt{?}'', if the language allows it.
In that case, the expression would also be of star-height~1.
\end{remark}

When expressions have no concatenation under star, \problemfont{Walk Membership} becomes tractable in combined complexity, as stated below.

\begin{restatable}{theorem}{cypherlikethm}
    \ltheorem{cypherlike-ptime-path-memb}
    \problemfont{Walk Membership} is in PTIME under binding-trail semantics when restricted to expressions with no concatenation under star. The same holds under simple-run semantics when queries are restricted to the Glushkov automata of such expressions.
\end{restatable}

Under binding-trail semantics, matching a starred subexpression with no concatenation in a given walk amounts to counting that the number of repetitions of each edge in the walk is less than the number of atoms of the expression with compatible label. Under simple-run semantics, the proof relies on a reduction to matchings in bipartite graphs \cite[Section~26.3]{CormenEtAl2009} and is detailed in Appendix~\ref{a:cypherlike}.

\subsection{Extensions of regular expressions}
\lsection{exte-rege-expr}
 
We briefly discuss how simple-run and binding-trail semantics behave with respect to some common extensions of RPQs.

\begin{description}
\item[One-or-more repetitions]
Many formalisms allow writing $R^+$ as a shorthand for $R\cdot R^*$. While expanding the notation is not suitable in our setting (the two $R$ subexpressions in $R\cdot R^*$ would then be matched independently, resulting in a different behaviour), treating $R^+$ as a new operator poses no particular problem: the definitions of both binding-trail semantics and Glushkov automaton extend naturally over $+$.

\item[Arbitrary repetitions] Some formalisms allow arbitrary repetitions, denoted by `${}^{\{n,m\}}$' with $n\in\N$ and~$m\in\N\cup\{\infty\}$: $R^{\{n,m\}}$ means that~$R$ may be repeated between~$n$ and~$m$ times. Once again, in our setting, expanding the notation would change the query. On the other hand, allowing it as a new operator would make Lemma~\ref{l:cove-spac} false. For instance, expression $R^{\{6,\infty\}}$ on the running database (Fig.~\ref{f:runn-data}) would yield tuple $(s,t)$ under walk semantics but not under binding-trail semantics. The impact of this operator on other complexity results is left for future work.

\item[Backward atoms]
Allowing backward atoms~$\overright{a}$ in expressions, as for instance in 2RPQs \cite{AnglesEtAl2017}, poses no particular problem: transitions of the automaton that are labelled with a backward atom are simply paired with reversed edges of the database in the run database.

\item[Any-directed atoms]
Cypher and GQL allow any-directed atoms $\overboth{a}$, that match edges labelled by~$a$ forward or backward. Under binding-trail semantics, it is quite easy to construct an example that makes Lemma~\ref{l:cove-spac} (and Prop.~\ref{p.bindingtrail-contains-shortest}) false.
For instance, consider the expression $R=\left(\big(\overleft{\rR}+\overleft{\rG}\big)\cdot\overboth{\rR}\cdot\big(\overleft{\rR}+\overleft{\rG}\big)\right)^*$
and the database from Figure~\ref{f:runn-data}.
Then, the walk~$w=c_1\anonarrow c_2 \anonarrow c_3 \anonarrow c_3 \anonrevarrow c_2 \anonarrow t$ contradicts Lemma~\ref{l:cove-spac}.
This issue can be circumvented by expanding $\overboth{a}$ into~$\overleft{a}+\overright{a}$ rather than treating it as a new operator, once again with some effect on the semantics of the query.

\end{description}

\section{Conclusion}
\label{s:comp-sema-comp}

Table~\ref{t:recap}, below, presents a summary of the computational complexity of popular semantics and compares them with run-based semantics. We also emphasize the following comparison points that do not appear in the table.
\begin{itemize}
    \item Table~\ref{t:recap} paints a very negative picture of trail semantics, which may seem contradictory with its widespread popularity in practice. 
    We believe that the main strength of trail semantics lies in that their output provides some kind of coverage of the space of matches, which, in turn, enables rich aggregation and post-processing. 
    Run-based semantics improves on this property by giving some guarantees on the coverage (Lemma~\ref{l:cove-spac}).
    \item \problemfont{Walk Membership} is a theoretical counterpart to \problemfont{Query Evaluation}: only the latter is implemented in real systems. Thus, we believe that run-based semantics offer a reasonable compromise by having tractable \problemfont{Tuple Membership}, tractable \problemfont{Query Evaluation} and intractable \problemfont{Walk Membership}.
    It is in our view better than the other way around, as in trail semantics.
    \item Some tasks are ill-defined or meaningless under homomorphism or shortest-walk semantics. Indeed, counting \emph{all} matching walks leads to unboundedness under homomorphism semantics, and returns an information of dubious value under shortest-walk: 0 or 1 most of the time, and the number of uncomparable minimal walks otherwise.
\end{itemize}

\begin{table}[ht] \centering
    \begin{tabular}{llllll}
        \toprule
        & Trail & Run-based &  Homomorphism & Shortest-walk 
        \\
        \midrule
        \problemfont{Tuple Membership} & Intractable & Tractable & Tractable & Tractable
        \\
        \problemfont{Tuple Multiplicity} & Intractable & Intractable & Ill-defined  & Meaningless
        \\
        \problemfont{Query Evaluation} & Intractable & Tractable & Ill-defined & Tractable
        \\
        \problemfont{Deduplicated Query Eval.} & Intractable & Open & Ill-defined & Open
        \\
        \problemfont{Walk Membership} & Tractable & Intractable & Tractable & Tractable
        \\
        \bottomrule
    \end{tabular}
    \caption{Summary of computational complexity}%
    \label{t:recap}%
\end{table}%

In conclusion, simple-run and binding-trail semantics provide good computational properties overall, supports bag semantics and rich aggregation.
However, while the RPQ formalism is a good model of the navigational part of most query languages over graph databases, it does not capture their ability to collect and compare data values along tested walks. Extending our proposed framework to handle data values lying in the edges and vertices of the database constitutes our main challenge going forward.


\bibliography{bibliography}

\input{appendix}

\end{document}

%% file: figure/tikzstyle.tex
\newlength{\minnoderadius}
\newlength{\shortenlength}
\newlength{\nodelinewidth}
\newlength{\arrowwidth}
\def\loopangle{24}
\tikzset{%
    bend angle=20,
    >={Stealth[width=6pt,length=6pt]},
    node/.style={circle, line width=\nodelinewidth, draw, black, inner sep=2pt, outer sep=.5*\nodelinewidth, minimum height=\minnoderadius, minimum width=\minnoderadius},
    vertex/.style={fill=black,inner sep=1.5pt, circle},
    state/.style={node},  
    run state/.style={draw,rounded rectangle},
    preedge/.style={-, draw, black, line width=1pt, rounded corners=5pt,pos=.4, shorten >=\shortenlength},
    edge/.style={preedge,->},
    redge/.style={preedge,<-},
    initialedge/.style={edge, shorten > =0pt, shorten < =0pt},
    finaledge/.style={redge, shorten > =0pt, shorten < =0pt},
    borderedge/.style={edge, -, color=white, line width=5pt, shorten >=\shortenlength-2pt, >={Stealth[width=12pt,length=12pt]}},
    vm loop/.style={edge, pos=.5, looseness = 8},    
    graph loop/.style={edge, pos=.5, looseness = 40, shorten >=4pt},    
    road/.style={edge, color=fred},
    hl/.style={edge, line width = 1.5mm, color=borange, shorten >=1pt},
    ferry/.style={edge,color=fpurple},
    gas/.style={edge,color=fblue},
    start/.style={edge,color=fblue},
    end/.style={edge,color=fblue},
    >=stealth,
    north west loop/.style={vm loop, in={\the\numexpr 135 + \loopangle\relax}, 
                                     out ={\the\numexpr 135 - \loopangle\relax}},
    north east loop/.style={vm loop, in={\the\numexpr 45 + \loopangle\relax}, 
                                     out ={\the\numexpr 45 - \loopangle\relax}},
    south west loop/.style={vm loop, in={\the\numexpr -135 + \loopangle\relax}, 
                                     out ={\the\numexpr -135 - \loopangle\relax}},
    south east loop/.style={vm loop, in={\the\numexpr -45 + \loopangle\relax}, 
                                     out ={\the\numexpr -45 - \loopangle\relax}},
    north loop/.style={vm loop, in={\the\numexpr 90 + \loopangle\relax}, 
                                out ={\the\numexpr 90 - \loopangle\relax}},
    south loop/.style={vm loop, in={\the\numexpr 270 - \loopangle\relax}, 
                                out ={\the\numexpr 270 + \loopangle\relax}},
    east loop/.style={vm loop, in={\the\numexpr 0 + \loopangle\relax}, 
                                out ={\the\numexpr 0 - \loopangle\relax}},
    west loop/.style={vm loop, in={\the\numexpr 180 - \loopangle\relax}, 
                                out ={\the\numexpr 180 + \loopangle\relax}},                
    node distance = \nodedist,
}

\newlength{\nodedist}
\newlength{\initfinaldist}
\newcommand{\initial}[2][180]{\path (#2.#1) ++(#1:\initfinaldist) coordinate (#2-initial-#1);
                              \path[initialedge] (#2-initial-#1) to (#2.#1);}

\newcommand{\final}[2][0]{\path (#2.#1) ++(#1:\initfinaldist) coordinate (#2-final-#1);
                          \path[initialedge] (#2.#1) to (#2-final-#1);}         
                          
\newcommand{\initialfinal}[2][0]{%
    \def\angleI{\the\numexpr #1 + 15 \relax}
    \def\angleII{\the\numexpr #1 - 15 \relax}
    \path (#2.\angleI) ++(#1:\initfinaldist) coordinate 
        (#2-initialfinal1-#1);
    \path[initialedge] (#2.\angleI) to         
        (#2-initialfinal1-#1);
    \path (#2.\angleII) ++(#1:\initfinaldist) coordinate     (#2-initialfinal2-#1);
    \path[finaledge] (#2.\angleII) to
      (#2-initialfinal2-#1);
}
\definecolor{vert}{rgb}{0,.55,0.20}
\definecolor{bleu}{rgb}{0,0,0.75}
\definecolor{rouge}{rgb}{.75,0,0}

%% file: macros.tex
\DeclareMathOperator{\tgt}{\mathop{\text{\normalfont\textsc{Tgt}}}}
\DeclareMathOperator{\src}{\mathop{\text{\normalfont\textsc{Src}}}}
\DeclareMathOperator{\lbl}{\mathop{\text{\normalfont\textsc{Lbl}}}}
\DeclareMathOperator{\len}{\mathop{\text{\normalfont\textsc{Len}}}}

\DeclareMathOperator{\lang}{\mathop{\text{\normalfont\textsc{L}}}}
\DeclareMathOperator{\card}{\mathop{\text{\normalfont\textsc{Card}}}}
\DeclareMathOperator{\trail}{\mathop{\text{\normalfont\textsc{Trail}}}}
\DeclareMathOperator{\simple}{\mathop{\text{\normalfont\textsc{Simple}}}}
\DeclareMathOperator{\bindingtrail}{\mathop{\text{\normalfont\textsc{BindingTrail}}}}
\DeclareMathOperator{\distinct}{\mathop{\text{\normalfont\textsc{Distinct}}}}

\DeclareMathOperator{\Endpoints}{\mathop{\text{\normalfont\textsc{Endpoints}}}}

\DeclareMathOperator{\im}{\mathop{\text{\normalfont\textsc{Im}}}}
\newcommand{\projD}{\mathop{\pi_D}}

\newcommand{\compose}{\mathbin{\circ}}

\newcommand{\sem}[2][N]{\llbracket #2\rrbracket_{#1}}
\newcommand{\match}[2]{\mathop{\text{\normalfont\textsc{Match}}_{#1}}(#2)}

\newcommand{\rarrow}[2][]{\xrightarrow[#1]{#2}}
\newcommand{\anonarrow}{\rightarrow}
\newcommand{\anonrevarrow}{\leftarrow}

\def\ilnode#1,{\if\detokenize{#1}@\else#1\expandafter\iledge\fi}
\def\iledge#1,{\if\detokenize{#1}@\else\xrightarrow{#1}\expandafter\ilnode\fi}

\newcommand{\statesof}[1]{Q_{#1}}
\newcommand{\transitionsof}[1]{\Delta_{#1}}
\newcommand{\alphabetof}[1]{\Sigma_{#1}}
\newcommand{\initialsof}[1]{I_{#1}}
\newcommand{\finalsof}[1]{F_{#1}}
\newcommand{\stdaut}[1]{%
#1=\aut{%
    \alphabetof{#1},%
    \statesof{#1},%
    \transitionsof{#1},%
    \initialsof{#1},%
    \finalsof{#1}%
}}

\newcommand{\Ac}{\mathcal{A}}
\newcommand{\Bc}{\mathcal{B}}

\newcommand{\comput}{\textsf{Comp}}

\newcommand{\aut}[1]{\langle#1\rangle}

\newcommand{\computof}[1]{\comput(#1)}

\newcommand{\finit}{\eta_i}
\newcommand{\ffinal}{\eta_f}

\newcommand{\N}{\mathbb{N}}

\makeatletter
\newcommand{\set}[2][]{\{#1{\kern1\nulldelimiterspace}#2{#1\kern1\nulldelimiterspace\}}}
\newcommand{\setst}{\@ifstar{\autosetst}{\paramsetst}}
\newcommand{\autosetst}[2]{\left\{\,#1\,\middle|\,#2\,\right\}}
\newcommand{\paramsetst}[3][]{#1\{\,#2\mathbin{#1|}#3{\,#1\}}}

\newcommand{\bag}{\@ifstar{\autobag}{\parambag}}
\newcommand{\autobag}[1]{\llbrace*#1\rrbrace*}
\newcommand{\parambag}[2][]{\llbrace[#1]#2\rrbrace[#1]}
\newcommand{\bagst}{\@ifstar{\autobagst}{\parambagst}}
\newcommand{\autobagst}[2]{\llbrace*\,#1\,\middle|\,#2\,\rrbrace*}
\newcommand{\parambagst}[3][]{\llbrace[#1]\,#2\mathbin{#1|}#3\,\rrbrace[#1]}

\newcommand{\llbrace}{\@ifstar{\leftllbrace}{\paramllbrace}}
\newcommand{\paramllbrace}[1][]{{#1\{\hspace*{-.25em}#1\{}}
\newcommand{\leftllbrace}{\left\lbrace\kern-3\nulldelimiterspace\middle\lbrace}
\newcommand{\rrbrace}{\@ifstar{\rightrrbrace}{\paramrrbrace}}
\newcommand{\paramrrbrace}[1][]{{#1\}}\hspace*{-.25em}{#1\}}}
\newcommand{\rightrrbrace}{\middle\rbrace\kern-3\nulldelimiterspace\right\rbrace}
\makeatother

\newcommand{\overleft}[1]{\overrightarrow{#1}}
\newcommand{\overright}[1]{\overleftarrow{#1}}
\newcommand{\overboth}[1]{\overleftrightarrow{#1}}

\newcommand{\claimqed}{}
\newenvironment{proofofclaim}{\begin{proof}[Proof of Claim~\theclaim{}]{\renewcommand{\qed}{}\popQED}\pushQED{\claimqed}}{\end{proof}}

\newcommand{\ldefinition}[1]{\label{d.#1}}

\newcommand{\llemma}[1]{\label{l.#1}}
\newcommand{\lproposition}[1]{\label{p.#1}}

\newcommand{\lremark}[1]{\label{r.#1}}

\newcommand{\lsection}[1]{\label{s.#1}}

\newcommand{\lfigure}[1]{\label{f.#1}}
\newcommand{\ltheorem}[1]{\label{t.#1}}
\newcommand{\lequation}[1]{\label{eq.#1}}

\newcommand{\preprocgenref}[2]{}
\newcommand{\generalref}[2]{%
  \preprocgenref{#1}{#2}%
  \ifthenelse{\equal{#1}{eq}}%
  {(\ref{#1.#2})}%
  {\ref{#1.#2}}%
}
\newcommand{\generalpageref}[2]{\pageref{#1.#2}}

\makeatletter
\newcommand*{\ralgorithm}{\@ifstar{\generalref{a}}{Algorithm~\ralgorithm*}}
\newcommand*{\palgorithm}{\@ifstar{\generalpageref{a}}{page~\palgorithm*}}

\newcommand*{\rcorollary}{\@ifstar{\generalref{c}}{Corollary~\rcorollary*}}
\newcommand*{\pcorollary}{\@ifstar{\generalpageref{c}}{page~\pcorollary*}}

\newcommand*{\rconjecture}{\@ifstar{\generalref{cj}}{Conjecture~\rconjecture*}}
\newcommand*{\pconjecture}{\@ifstar{\generalpageref{cj}}{page~\pconjecture*}}

\newcommand*{\rdefinition}{\@ifstar{\generalref{d}}{Definition~\rdefinition*}}
\newcommand*{\pdefinition}{\@ifstar{\generalpageref{d}}{page~\pdefinition*}}

\newcommand*{\rexample}{\@ifstar{\generalref{e}}{Example~\rexample*}}
\newcommand*{\pexample}{\@ifstar{\generalpageref{e}}{page~\pexample*}}

\newcommand*{\rlemma}{\@ifstar{\generalref{l}}{Lemma~\rlemma*}}
\newcommand*{\plemma}{\@ifstar{\generalpageref{l}}{page~\plemma*}}

\newcommand*{\rproposition}{\@ifstar{\generalref{p}}{Proposition~\rproposition*}}
\newcommand*{\pproposition}{\@ifstar{\generalpageref{p}}{page~\pproposition*}}

\newcommand*{\rproperty}{\@ifstar{\generalref{pp}}{Property~\rproperty*}}
\newcommand*{\pproperty}{\@ifstar{\generalpageref{pp}}{page~\pproperty*}}

\newcommand*{\rprocedure}{\@ifstar{\generalref{pc}}{Procedure~\rprocedure*}}
\newcommand*{\pprocedure}{\@ifstar{\generalpageref{pc}}{page~\pprocedure*}}

\newcommand*{\rremark}{\@ifstar{\generalref{r}}{Remark~\rremark*}}
\newcommand*{\premark}{\@ifstar{\generalpageref{r}}{page~\premark*}}

\newcommand*{\rnotation}{\@ifstar{\generalref{n}}{Notation~\rnotation*}}
\newcommand*{\pnotation}{\@ifstar{\generalpageref{n}}{page~\pnotation*}}

\newcommand*{\rsection}{\@ifstar{\generalref{s}}{Section~\rsection*}}
\newcommand*{\psection}{\@ifstar{\generalpageref{s}}{page~\psection*}}

\newcommand*{\rtable}{\@ifstar{\generalref{t}}{Table~\rtable*}}
\newcommand*{\ptable}{\@ifstar{\generalpageref{t}}{page~\ptable*}}

\newcommand*{\rfigure}{\@ifstar{\generalref{f}}{Figure~\rfigure*}}
\newcommand*{\pfigure}{\@ifstar{\generalpageref{f}}{page~\pfigure*}}

\newcommand*{\requation}{\@ifstar{\generalref{eq}}{Equation~\requation*}}
\newcommand*{\pequation}{\@ifstar{\generalpageref{eq}}{page~\pequation*}}

\newcommand*{\rtheorem}{\@ifstar{\generalref{t}}{Theorem~\rtheorem*}}
\newcommand*{\ptheorem}{\@ifstar{\generalpageref{t}}{page~\ptheorem*}}

\newcommand*{\rclaim}{\@ifstar{\generalref{cl}}{Claim~\rclaim*}}
\newcommand*{\pclaim}{\@ifstar{\generalpageref{cl}}{page~\pclaim*}}

\newcommand*{\rfact}{\@ifstar{\generalref{f}}{Fact~\rclaim*}}
\newcommand*{\pfact}{\@ifstar{\generalpageref{f}}{page~\pclaim*}}
\makeatother

\usepackage[framemethod=TikZ]{mdframed}
\newlength{\problemmargin}
\newlength{\problempadding}
\newlength{\problemsep}
\newlength{\problemtitlepadding}
\newcommand{\problembullet}{\textbullet~~}
\newcommand{\problemfont}[1]{\textsc{#1}}
\makeatletter
\newenvironment{problem}[1]{
    
    \addvspace{.5\baselineskip\@plus.2\baselineskip\@minus.2\baselineskip}
    
    \begin{mdframed}[%
        frametitle={\hspace*{\dimexpr\problemtitlepadding-\problempadding\relax}#1},
        frametitlefont=\large\sc,
        innerleftmargin=\problempadding,
        innerrightmargin=\problempadding,
        innerbottommargin=\problempadding,
        innertopmargin=\problempadding,
        frametitleaboveskip=\problemtitlepadding,
        frametitlebelowskip=\problemtitlepadding,
        leftmargin=\problemmargin,
        rightmargin=\problemmargin,
        frametitlerule=true,
        frametitlebackgroundcolor=black!10,
        nobreak,
        roundcorner=0pt,
        linewidth=0.5pt,
        frametitlerulewidth=0.5pt,
        ]%
    \description[topsep=0pt,itemsep=\problemsep,labelindent=0cm,leftmargin=\widthof{\problembullet},font=~\textbullet~\normalfont]%
}{%
    \enddescription\end{mdframed}
    
    \addvspace{.5\baselineskip\@plus.2\baselineskip\@minus.2\baselineskip}
}
\makeatother

\usepackage{listings, bold-extra}
\newcommand{\ttpcr}{\renewcommand{\ttdefault}{pcr}\ttfamily}
\makeatletter
\lstnewenvironment{cypher}[1][]{%
    \noindent\minipage[b]{\linewidth}%
    \centering%
    \tabular{@{}c@{}}%
}{\endtabular\endminipage\vspace{5pt}}
\makeatother
\makeatletter
\lstnewenvironment{gql}[1][]{%
    \noindent\minipage[b]{\linewidth}%
    \centering%
    \tabular{@{}c@{}}%
}{\endtabular\endminipage\vspace{5pt}}
\makeatother

\newcommand{\gl}{Gl}

\ifdraft
    \newcommand{\mcomment}[2]{{\color{blue}\textbf{(#1)}}\footnote{\textbf{#1:} #2}}%
\else
    \newcommand{\mcomment}[2]{}%
\fi

\newcommand{\cstack}[1]{\begin{array}{@{}c@{}}#1\end{array}}
\newcommand{\bistack}[2]{\left\langle\cstack{#1\\#2}\right\rangle}

\def\gasL{\textbf{G}\footnotesize as}
\def\roadL{\textbf{R}\footnotesize oad}
\def\ferryL{\textbf{F}\footnotesize erry}

\newcommand{\rR}{\mathbf{R}}
\newcommand{\rG}{\mathbf{G}}
\newcommand{\rF}{\mathbf{F}}

%% file: figure/roads.tex
\input{figure/roads-tikz-style}%
\begin{tikzpicture}[node distance = .75\linewidth]
\newcommand{\dist}{22mm}

\node[vertex] (start) {};

\path (start) 
    ++(0:\dist) node[vertex] (1) {} 
    ++(0:\dist) node[vertex] (2) {} 
    ++(0:\dist) node[vertex] (end) {};
    
\path (1)  ++(-60:\dist) node[vertex] (3) {};


\path (start) ++(90:1pt) node[anchor=south east] (startlabel) {\small$s$};
\path (end) ++(90:1pt) node[anchor=south west] (endlabel) {\small$t$};

\path (1) ++(90:1pt) node[anchor=south] (2label) {\small$c_1$};

\path (2) ++(90:1pt) node[anchor=south] (3label) {\small$c_2$};

\path (3) ++(180:3pt) node[anchor=east] (4label) {\small$c_3$};

\path[road] (start) to node[pos=.45,below] {\roadL} (1);

\path[road] (1) to node[pos=.45,below] {\roadL} (2); 
\path[road] (2) to node[pos=.45,below,sloped] {\roadL} (3); 
\path[road] (3) to node[pos=.45,below,sloped] {\roadL} (1); 
\path[road] (2) to node[pos=.45,below] {\roadL} (end);


\path[ferry, rounded corners] 
    let \p1=(end) in
    (start) -- ++(0,7mm) 
        to node[pos=.5,above] {\ferryL} 
    (\x1,\y1+7mm) -- (end);


\path[south loop,graph loop,gas] (3) to node[pos=.3,anchor=west] (gaslabel) {\gasL} (3);




\end{tikzpicture}%

%% file: figure/a.tex
\input{figure/roads-tikz-style}%
\begin{tikzpicture}
\setlength{\nodedist}{18mm}  
\node[state] (i) {$0$};
\initial[180]{i}
\node[state, below of = i] (f) {$1$};
\final{f}
\path[edge] (i) to[pos=.4] node[right] {\gasL} (f);

\path[south loop] (f) to node[below] 
{\begin{tabular}{c}
 \ferryL,\\
 \roadL\end{tabular}} (f);
 
\path[north loop] (i) to node[above] 
    {\begin{tabular}{c}
     \ferryL,\\
     \roadL\end{tabular}} (i);

\end{tikzpicture}%

%% file: figure/roads-product.tex
\input{figure/roads-tikz-style}%
\begin{tikzpicture}[node distance = .75\linewidth]%
\newcommand{\dist}{22.8mm}%

 \begin{scope}
    \node[run state] (start-left) {$s,0$};
    
    \path (start-left) 
        ++(0:\dist) node[run state] (1-left) {$c_1,0$} 
        ++(0:\dist) node[run state] (2-left) {$c_2,0$} 
        ++(0:\dist) node[run state, dashed] (end-left) {$t,0$};
        
    \path (1-left)  ++(-60:\dist) node[run state] (3-left) {$c_3,0$};
    
    \path[road] (start-left) to node[pos=.45,below] {\roadL} (1-left);
    
    \path[road] (1-left) to node[pos=.45,below] {\roadL} (2-left); 
    \path[road] (2-left) to node[pos=.45,below,sloped] {\roadL} (3-left); 
    \path[road] (3-left) to node[pos=.45,below,sloped] {\roadL} (1-left); 
    \path[road,dashed] (2-left) to node[pos=.45,below] {\roadL} (end-left);
    
    \path[ferry, rounded corners,dashed] 
        let \p1=(end-left) in
        (start-left) -- ++(0,7mm) 
            to node[pos=.5,above] {\ferryL} 
        (\x1,\y1+7mm) -- (end-left);
    
\end{scope}

\begin{scope}[xshift=1.25*\dist,yshift=-1.73*\dist]
    \node[run state,dashed] (start-right) {$s,1$};
    
    \path (start-right) 
        ++(0:\dist) node[run state] (1-right) {$c_1,1$} 
        ++(0:\dist) node[run state] (2-right) {$c_2,1$} 
        ++(0:\dist) node[run state] (end-right) {$t,1$};
        
    \path (1-right)  ++(60:\dist) node[run state] (3-right) {$c_3,1$};
    
    
    
    

    \path[road,dashed] (start-right) to node[pos=.45,above] {\roadL} (1-right);
    
    \path[road] (1-right) to node[pos=.45,above] {\roadL} (2-right); 
    \path[road] (2-right) to node[pos=.45,above,sloped] {\roadL} (3-right); 
    \path[road] (3-right) to node[pos=.45,above,sloped] {\roadL} (1-right); 
    \path[road] (2-right) to node[pos=.45,above] {\roadL} (end-right);
    
    \path[ferry, rounded corners, dashed] 
        let \p1=(end-right) in
        (start-right) -- ++(0,-7mm) 
            to node[pos=.5,below] (highwaylabel-left) {\ferryL} 
        (\x1,\y1-7mm) -- (end-right);
\end{scope}

\path[gas] (3-left) to node[pos=.45, below, sloped] (gaslabel) {\gasL} (3-right);

\path
    let \p1=(current bounding box.east) ,
        \p2=(current bounding box.north) 
    in (\x1,\y2) coordinate (bbne);
\path
    let \p3=(current bounding box.west),
        \p4=(highwaylabel-left.base)
    in (\x3,\y4) coordinate (bbsw);
\pgfresetboundingbox;
\path[use as bounding box] (bbne) rectangle (bbsw);

\end{tikzpicture}%

%% file: figure/grid.tex
\begin{tikzpicture}%

\tikzset{
    varedge/.style={edge,color=blue!75!black},
    recalledge/.style={edge,color=blue!75!black},
    keepedge/.style={edge,color=green!50!black},
    evaledge/.style={edge,color=red!75!black},
    checkedge/.style={edge,color=brown!75!black},
    invertedge/.style={edge,color=orange!75!black},
    resetedge/.style={edge,color=black},
    edge1/.style={font=\small},
    edge2/.style={font=\small},
    edge3/.style={font=\small},
}%

\input{figure/grid_nodes}%
\input{figure/grid_edges2}%
\input{figure/grid_edges3}%


\end{tikzpicture}

%% file: figure/asapbbs.tex
\begin{tikzpicture}
\setlength{\nodedist}{18mm}  
\node[node] (i) {\texttt{S}};
\node[node, right of = i] (f) {\texttt{T}};

\path[edge] (i) to node[above] {$a,b$} (f);

\path[east loop] (f) to node[right] {$b$} (f);
\path[west loop] (i) to node[left] {$a$} (i);
\end{tikzpicture}%

%% file: appendix.tex
\appendix
\newcounter{appendix}\setcounter{appendix}{0}
\renewcommand{\theappendix}{\Alph{appendix}}
\renewcommand{\thesection}{\theappendix\arabic{section}}
\renewcommand{\thetheorem}{\theappendix\arabic{theorem}}
\renewcommand{\theproposition}{\theappendix\arabic{proposition}}
\renewcommand{\thecorollary}{\theappendix\arabic{corollary}}
\renewcommand{\thelemma}{\theappendix\arabic{lemma}}
\renewcommand{\thedefinition}{\theappendix\arabic{definition}}

\newcommand{\newappendix}[1]{%
  \refstepcounter{appendix}%
  \setcounter{section}{0}%
  \clearpage%
  {%
    \noindent\huge\bf Appendix \theappendix:\quad
    #1%
    \vspace*{10mm}
  }
  
}

\ifdraft
    \input{appendix/notes}
\fi

\input{appendix/recap}
\input{appendix/walk-membership}
\input{appendix/topo_coding}

\input{appendix/proof_cypherlike}

\ifdraft
    \input{appendix/trailrun}
\fi

%% file: appendix/notes.tex
\newappendix{Notes}
\section*{Notes}
\markboth{Notes}{Notes}
\subsection*{Reference à lire}

\begin{itemize}
    \item Reference intéressante: \cite{BonifatiEtAl2021}. Ils présentent une autre approche pour résoudre le type de problème qu'on traite ici. 
    Au lieu de compter le nombre de résultats, on compte jusqu'à un certain seuil. Ceci permettrait de réduire la complexité du comptage du nombre de solutions.  Elle a l'air d'être compatible avec notre vue. A vérifier.
    \item Référence possiblemente pertinente: \cite{CaronZiadi2000}.  Ils caractérisent la structure d'un automate de Glushkov. A vérifier le rapport avec Prop~\ref{p.topo-codi} et consoeurs.
\end{itemize}

\section*{Titre?}
\begin{itemize}
    \item Simple-run semantics
    \item Simple-run semantics, theory and practice meet halfway.
    \item Simple-run semantics for RPQs
    \item Simple-run, new semantics for regular path queries based on practictal observation
    \item Run-based semantics
    \item Run-based semantics for RPQs
    \item Defining RPQ semantics based on run
\end{itemize}

%% file: appendix/recap.tex
\newappendix{Recap figure}

\noindent%
\begin{minipage}{\textwidth}
    \centering
    \input{figure/recap.tex}
    
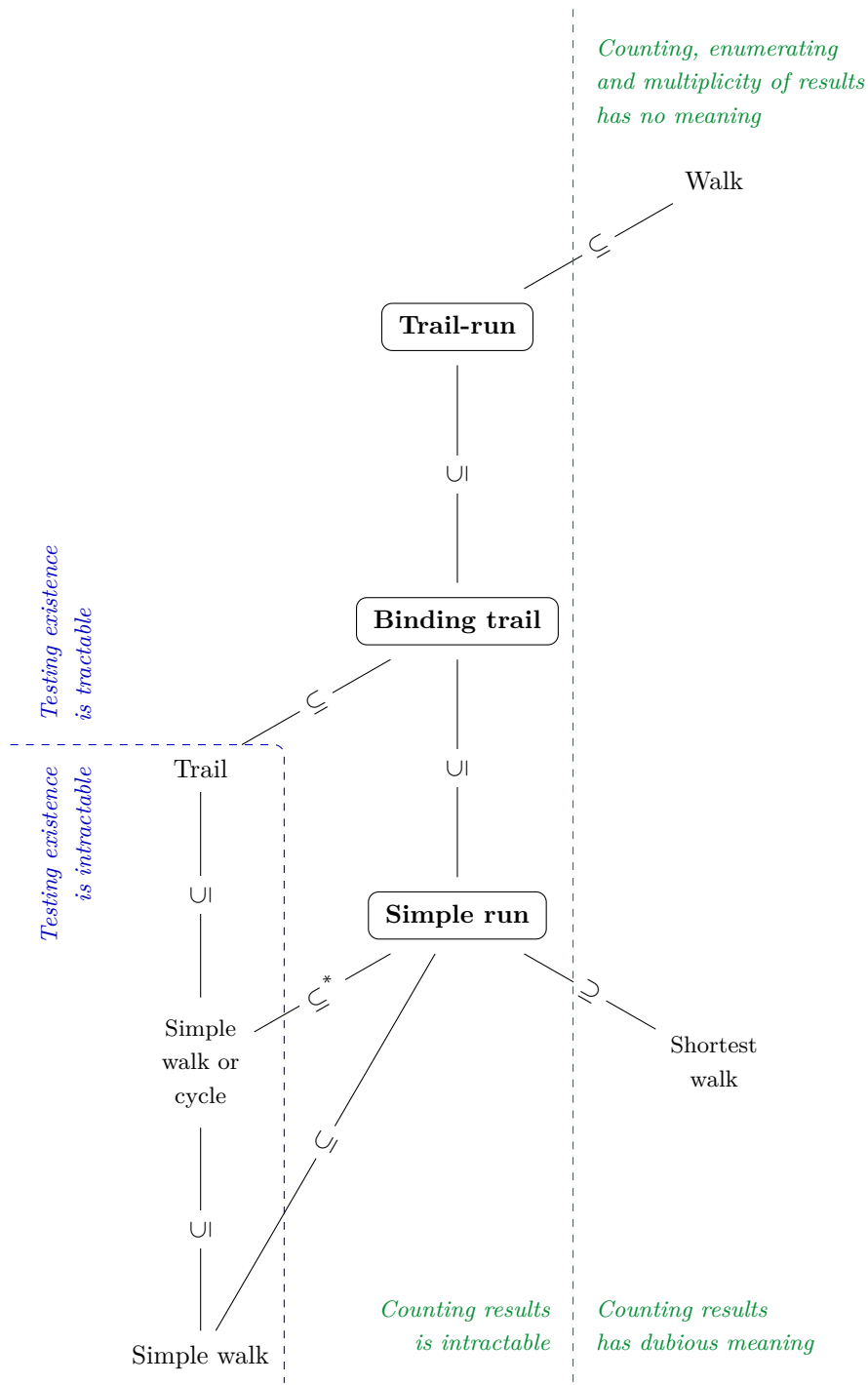
\captionof{figure}{Summary of the different semantics mentioned in the introduction. (The inclusion marked with * holds only if the automaton is standard.)}
    \lfigure{sema-incl}
\end{minipage}

%% file: figure/recap.tex
\begin{tikzpicture}
    \newcommand{\semname}[1]{\rule{0pt}{1.2ex}\smash{#1}}
    \newlength{\vmh}\setlength{\vmh}{40mm}
    \tikzstyle{sem}=[
                     ellipse,
                     outer sep=2mm, 
                     inner sep=0pt,
                     rectangle, 
                     rounded corners, 
                    ]
    \tikzstyle{oursem}=[font=\bf,
                        sem,
                        draw, 
                        inner sep=1.5ex
                       ]
    \tikzstyle{incedge}=[draw,
                        ]
    \tikzstyle{classlabel}=[inner sep=2ex]
                    
    \node[sem] (simple-walk) at (0,0) {Simple walk} ;
    \path (simple-walk) 
    ++(90:\vmh) node[sem] (simple-walk-cycle){\begin{tabular}{@{}c@{}}Simple \\walk or\\cycle\end{tabular}}
    ++(90:\vmh) node[sem] (trail) {Trail};
    
    \path (simple-walk-cycle) 
    ++(30:\vmh) node[oursem] (simple-run){\semname{Simple run}}
    ++(90:\vmh) node[oursem] (binding-trail){\semname{Binding trail}}
    ++(90:\vmh) node[oursem] (trail-run){\semname{Trail-run}} ;
        
    \path (simple-run) 
    ++(-30:\vmh) node[sem] (shortest-walk){\begin{tabular}{@{}c@{}}Shortest\\walk\end{tabular}};
    
    \path (trail-run) 
    ++(30:\vmh) node[sem] (walk) {Walk};
    
    \newcommand{\inclusion}[3][$\subseteq$]{\path[incedge] (#2) to node[sloped,fill=white] {#1} (#3);}
    
    \inclusion{simple-walk}{simple-walk-cycle}
    \inclusion{simple-walk-cycle}{trail}
    
    \inclusion[${\subseteq}^{\,\text{*}}$]{simple-walk-cycle}{simple-run};
    \inclusion{trail}{binding-trail};
    
    \inclusion{simple-run}{binding-trail}
    \inclusion{binding-trail}{trail-run}
    \inclusion[$\supseteq$]{shortest-walk}{simple-run}
    \inclusion{simple-walk}{simple-run}
    
    \inclusion{trail-run}{walk}

    \path (trail.north west) ++(180:.5\vmh) coordinate (untract-1);
    \path let \p1 = (simple-walk.east) in 
          let \p2 = (trail.north) in 
          coordinate (untract-2) at (\x1,\y2);
    \path let \p1 = (simple-walk.east) in 
          let \p2 = (simple-walk.south east) in 
          coordinate (untract-3) at (\x1,\y2);
    
    \draw[color=bleu, dashed,rounded corners] 
        (untract-1) -- (untract-2) -- (untract-3);
    

    \node[classlabel,color=bleu,anchor=north east,rotate=90] (untract) at (untract-1) {\it\begin{tabular}{@{}r@{}}Testing existence\\is intractable\end{tabular}};
    \node[classlabel,color=bleu,anchor=north west,rotate=90] (untract) at (untract-1) {\it\begin{tabular}{@{}l@{}}Testing existence\\is tractable\end{tabular}};

    
    \path (shortest-walk.north west) ++(150:0\vmh) coordinate (count-2);
    \path (shortest-walk.south west) ++(-90:.5\vmh) coordinate (count-3);


    \path let \p1 = (simple-walk.south) in 
          let \p2 = (binding-trail.east) in 
          coordinate (enum-2) at (\x2,\y1);
    \path let \p1 = (walk.north) in 
          let \p2 = (binding-trail.east) in 
          (\x2,\y1) ++(90:.5\vmh)
          coordinate (enum-1);
    \draw[color=vert,dashed,rounded corners] (enum-1) -- (enum-2);
    
    \node[classlabel,anchor=south east,color=vert] (count) at (enum-2) {\it\begin{tabular}{@{}r@{}}Counting results\\ is intractable\end{tabular}};
    \node[classlabel,anchor=south west,color=vert] (count) at (enum-2) {\it\begin{tabular}{@{}l@{}}Counting results\\ has dubious meaning\end{tabular}};
    \node[classlabel,anchor=north west,color=vert] (count) at (enum-1) {\it\begin{tabular}{@{}l@{}}Counting, enumerating\\ and multiplicity of results\\ has no meaning\end{tabular}};
\end{tikzpicture}

%% file: appendix/walk-membership.tex
\newappendix{Proof of \protect\rtheorem{walk-memb-untr}\label{a:walk-memb}}%

\newcommand{\negvarset}{\bar{X}}%
\newcommand{\posvarset}{X}%
\newcommand{\anyvarset}{X\cup\bar{X}}%
\newcommand{\negvar}[1]{\bar{#1}}%
\newcommand{\posvar}[1]{#1}%
\newcommand{\anyvar}[1]{\tilde{#1}}%
\newcommand{\powerset}{\mathbb{P}}%


\begin{falsestatement}{\rtheorem{walk-memb-untr}}
    \walkmembuntrstatement
\end{falsestatement}

The problem is obviously in NP as one can guess a run~$r$ in~$D\times\Ac$ and check that ${\projD(r)=w}$.
The proof of hardness is done by reduction from \problemfont{3-Sat}. We define a fixed automaton~$\Ac$ such that for any \problemfont{3-Sat} instance~$I$ we can build a polynomial size database~$D_I$ and walk~$p_I$ such that $I$ is satisfiable iff $p_I\in \sem[SR]{\Ac}(D_I)$.


\paragraph*{Preliminary warning.} Let us warn the reader that the sketch of proof as well as the \rfigure{reduc-graph}, \pfigure{reduc-graph} give a simplified version of the construction presented here.
The database $D_I$ defined in this formal proof has many more \emph{useless} vertices and edges in order to make the definition and statements easier to formulate. We omit those in the body of the paper to simplify \rfigure{reduc-graph} and give better intuition in the sketch of proof. \rfigure{DI-example} gives the database corresponding to the example of \rfigure{reduc-graph} for the encoding defined below.

\begin{figure}[h]%
 \includegraphics[width=\linewidth]{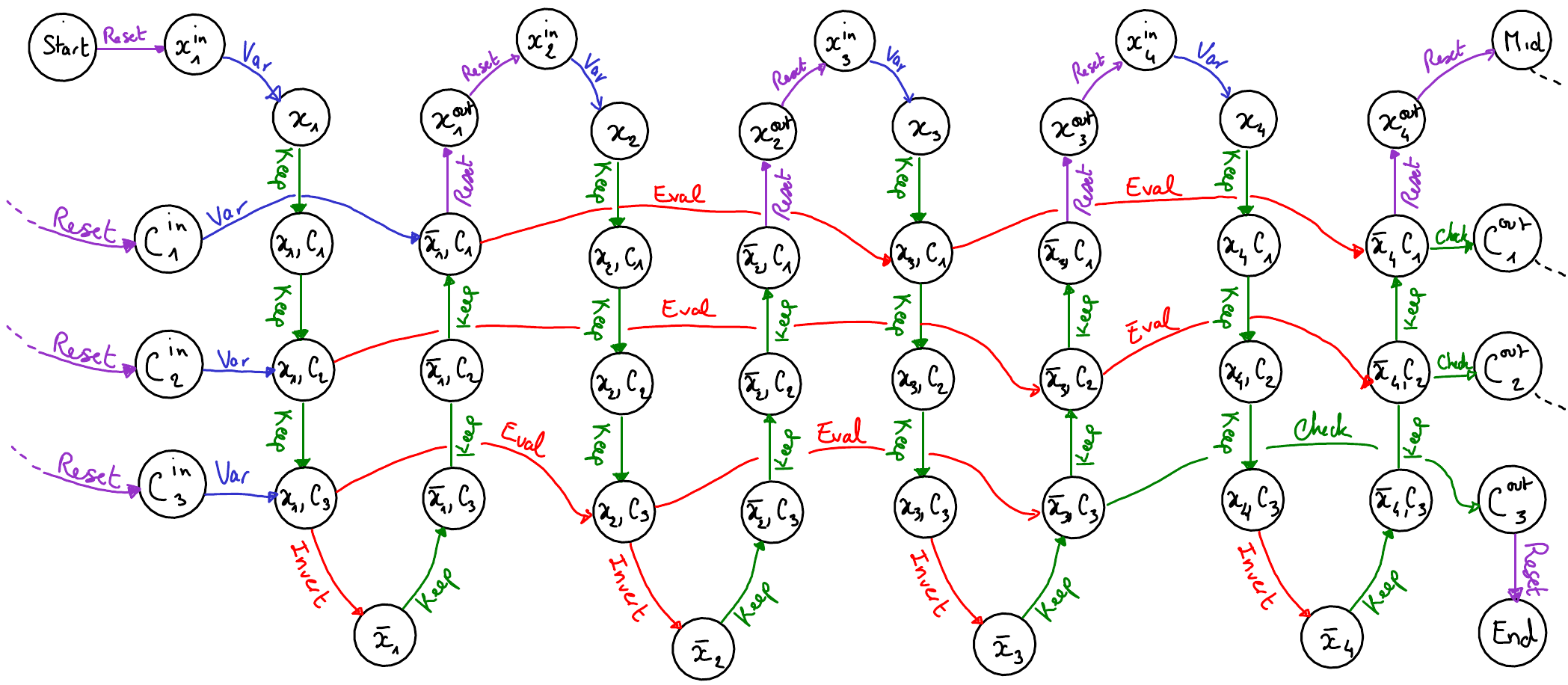}
 \caption{Drawing of the database $D_I$ encoding the instance $I=C_1\wedge C_2 \wedge C_3$ with ${C_1=\neg x_1 \vee x_3 \vee \neg x_4}$, 
    ${C_2=x_1\vee \neg x_3 \vee \neg x_4}$
    and ${C_3=x_1\vee x_2 \vee \neg x_3}$}

    \lfigure{DI-example}
\end{figure}

\section{The automaton~$\Ac$}
    Let $\Ac$ be the automaton defined as follow: 
    \begin{itemize}
        \item the alphabet is $\set{\checkL,\evalL,\invertL,\keepL,\resetL,\varL}$;
        \item $\Ac$ has three states $0$, $1$ and  $\top$;
        \item $\top$ is both the only intial state and the only final state;
        \item its transition table is given below. (See \rfigure{reduc-auto} for a graphical presentation.)
    \end{itemize}
\noindent{\setlength{\tabcolsep}{3pt}\def\arraystretch{1}%
    %
    \begin{tabular}[t]{lrcl}
    \toprule
        \keepL:   & $0$    & $\rightarrow$ & $0$ \\
          & $1$    & $\rightarrow$ & $1$ \\
          & $\top$ & $\rightarrow$ & $\top$ \\
    \midrule
        \varL:    & $\{0,1,\top\}$ & $\rightarrow$ & $\{0,1\}$ \\
    \bottomrule
    \end{tabular}
    \hspace*{0cm plus 2 fill}
    \begin{tabular}[t]{lrcl}
    \toprule
        \invertL: & $0$    & $\rightarrow$ & $1$ \\ 
          & $1$    & $\rightarrow$ & $0$ \\
    \midrule
        \resetL:  & $\{0,1,\top\}$ & $\rightarrow$ & $\top$\\
    \bottomrule
    \end{tabular}
    \hspace*{0cm plus 2 fill}
    \begin{tabular}[t]{lrcl}
    \toprule  
        \evalL:   & $1$        & $\rightarrow$ & $\{0,1\}$\\
                  & $\{0,\top\}$ & $\rightarrow$ & $\top$\\      
    \midrule
        \checkL:  & $\{0,\top\}$ & $\rightarrow$ & $\top$\\ 
    \bottomrule
    \end{tabular}
    %
}
    
\begin{figure*}[t]
    \newcommand{\separate}{%
        \hspace*{0cm plus 1 fil}\allowbreak\hspace*{0cm plus 1 fil}}%
    \begin{subfigure}[t]{.3\linewidth}%
        \centering%
        \fbox{\input{figure/reduc-auto-var}}
        \caption{Transitions labelled by $\varL$}
        \lfigure{reduc-auto-var}%
    \end{subfigure}
    \hfill
    \begin{subfigure}[t]{.3\linewidth}%
        \centering%
        \fbox{\input{figure/reduc-auto-keep}}
        \caption{Transitions labelled by $\varL$}
        \lfigure{reduc-auto-keep}
    \end{subfigure}
    \hfill
    \begin{subfigure}[t]{.3\linewidth}%
        \centering%
        \fbox{\input{figure/reduc-auto-invert}}
        \caption{Transitions labelled by $\varL$}
        \lfigure{reduc-auto-invert}
    \end{subfigure}
    
    \vspace*{10pt}
    
    \begin{subfigure}[t]{.3\linewidth}%
        \centering%
        \fbox{\input{figure/reduc-auto-eval}}
        \caption{Transitions labelled by $\varL$}
        \lfigure{reduc-auto-eval}
    \end{subfigure}
    \hfill
    \begin{subfigure}[t]{.3\linewidth}%
        \centering%
        \fbox{\input{figure/reduc-auto-check}}
        \caption{Transitions labelled by $\varL$}
        \lfigure{reduc-auto-check}
    \end{subfigure}
    \hfill
    \begin{subfigure}[t]{.3\linewidth}%
        \centering%
        \fbox{\input{figure/reduc-auto-reset}}
        \caption{Transitions labelled by $\varL$}
        \lfigure{reduc-auto-reset}
    \end{subfigure}
    \caption{Transitions of the automaton~$\Ac$ ($\top$ is initial and final)}
    \lfigure{reduc-auto}
\end{figure*}
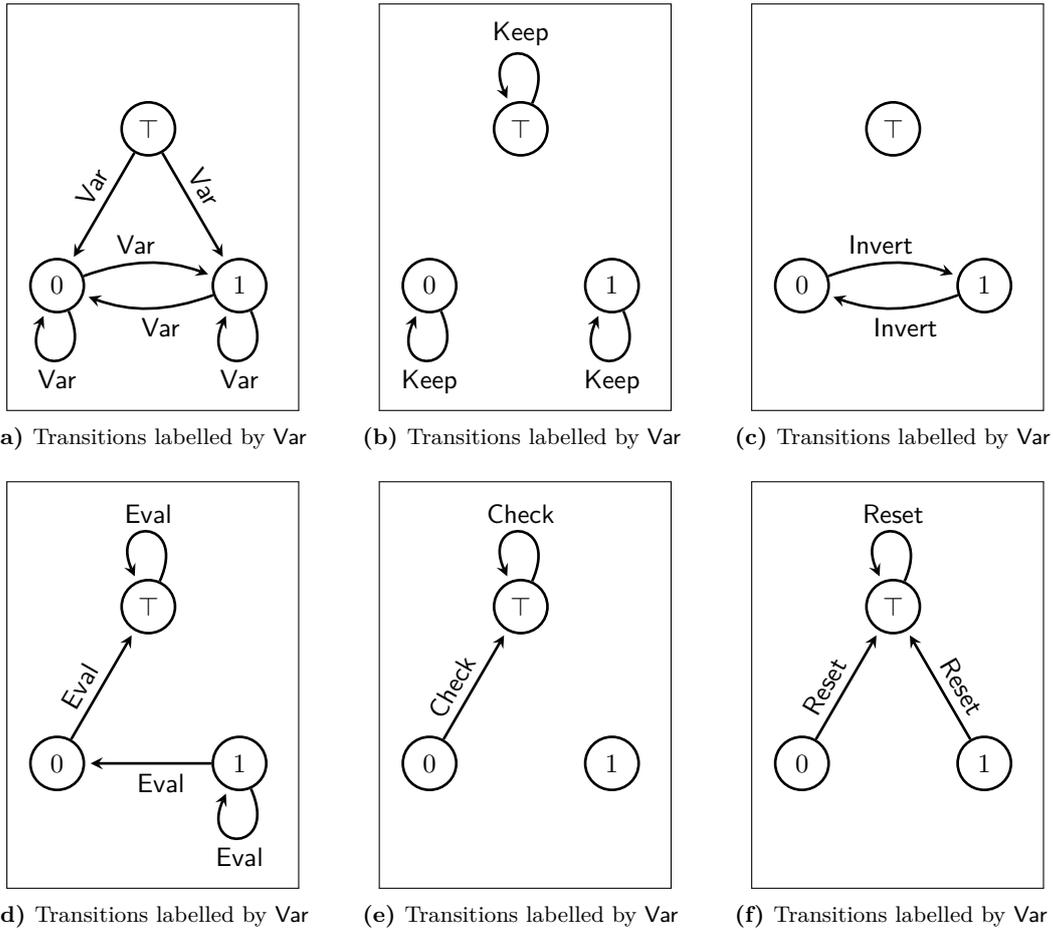
\section{SAT instance $I$}
Given a \problemfont{3-Sat} instance $I=C_1\wedge \cdots\wedge C_\gamma$, we denote by $\posvarset$ the set of the distinct variables~$\posvar{x}_1,\ldots,\posvar{x}_n$ appearing in~$I$ and by~$\negvarset$ the set of the corresponding negated version of these variables: $\negvar{x}_1,\ldots,\negvar{x}_n$.

In the following, we will generally use the notation~$\posvar{x}$, $\negvar{x}$ and $\anyvar{x}$ for elements in~$\posvarset$, $\negvarset$ and $\anyvarset$, respectively.
Typically we write $C_i=\anyvar{x}_\ell\vee\anyvar{x}_k\vee\anyvar{x}_m$.

\section{The Database~$D_I$}
\lsection{db-B}

Let $I$ be a \problemfont{3-Sat} instance. We can now build a database $D_I$ from~$I$. An example is given in \rfigure{DI-example}.

\subsection*{Vertices}

\begin{itemize}
    \item $D_I$ contains three special vertices: $\text{Start}$, $\text{Mid}$, and $\text{End}$
    \item For each~$\posvar{x}$ in~$\posvarset$, $D_I$ contains three vertices:~$\posvar{x}^{\text{in}}$, $\posvar{x}$, and $\posvar{x}^{\text{out}}$.
    \item For each $\negvar{x}$ in~$\negvarset$, $D$ contains one vertex:~$\negvar{x}$.
    \item For each element~$\anyvar{x}$ in~$\anyvarset$ and each clause~$C_i$, $D_I$ contains one vertex~$(\anyvar{x},C_i)$
    \item For each clause~$C_i$ we create two vertices: $C_i^\text{in}$ and~$C_i^\text{out}$.
\end{itemize}

\subsection*{Edges}
\lsection{walk-memb-untr-edge}

For each variable~$\posvar{x}\in\posvarset$ we add the edges from  the following walk~$p_{\posvar{x}}$ to~$D_I$.
    \begin{multline*}
           p_x = x^\text{in} \rarrow{\varL} x \rarrow{\keepL}(x,C_0)\rarrow{\keepL}\cdots \rarrow{\keepL}(x,C_\gamma)
           \\
           \rarrow{\invertL}\negvar{x}\rarrow{\keepL}(\negvar{x},C_\gamma)\rarrow{\keepL}\cdots \rarrow{\keepL}(\negvar{x},C_0)\rarrow{\resetL}x^\text{out}
    \end{multline*}
For each clause~$C_i=\anyvar{x}_k\vee\anyvar{x}_\ell\vee\anyvar{x}_m$, we add edges from the following walk~$p_{C_i}$ to~$D_I$.
    \begin{equation*}
        p_{C_i} = C_i^\text{in} \rarrow{\varL} (\anyvar{x}_k,C_i) \rarrow{\evalL} (\anyvar{x}_\ell,C_i) \rarrow{\evalL} (\anyvar{x}_m,C_i) \rarrow{\checkL} C_i^\text{out}
    \end{equation*}
Then we connect these walks by~$\resetL$-labelled edges as follows.
    \begin{align*}
    &\text{Start} \rarrow{\resetL} x_1^\text{in} \\
    \forall i,~1\leq i < n\quad &x_i^{\text{out}}\rarrow{\resetL}x_{i+1}^{\text{in}}\\
    &x_n^{\text{out}} \rarrow{\resetL} \text{Mid} \\[10pt]
    &\text{Mid} \rarrow{\resetL} C_1^{\text{in}} \\
    \forall i, 1\leq i < \gamma \quad &C_i^{\text{out}}\rarrow{\resetL}C_{i+1}^{\text{in}}\\
    & C_\gamma\rarrow{\resetL} \text{End}
\end{align*}

\section{The walk~$p_I$, and its components~$p_\text{setval}$ and~$p_\text{checksat}$}
    The walk~$p_I$ consists in traversal from Start to End going through every single edge of the database $D_I$. Intuitively, a first walk~$p_\text{setval}$ going through every $p_{x}$ will be used to define a valuation of the variables. A second walk~$p_\text{checksat}$ going through every $p_{C_i}$ will be used to check that this valuation makes the instance true. Formally, we define the walk~$p_I$ as~$p_\text{setval}\cdot p_\text{checksat}$ where ~$p_\text{setval}$ and~$p_\text{checksat}$ are as follows :
    \begin{align*}
                p_\text{setval} ={}&
        \\ 
        \multispan2{\makebox[\linewidth][r]{$\text{Start}\rarrow{\resetL} p_{x_1} \rarrow{\resetL} p_{x_2} \rarrow{\resetL} \cdots \rarrow{\resetL} p_{x_n} \rarrow{\resetL} \text{Mid}$}}
        \\
        p_\text{checksat} = {}& \\
        \multispan2{\makebox[\linewidth][r]{$\text{Mid}\rarrow{\resetL} p_{C_1}\rarrow{\resetL}p_{C_2}\rarrow{\resetL}\cdots\rarrow{\resetL}p_{C_\gamma}\rarrow{\resetL} \text{End}$}}
    \end{align*}
    Note that the equation above makes a slight abuse of notation: the $p_{x_i}$'s and $p_{C_i}$'s are walks instead of vertices: $p \rarrow{\resetL} p'$ means that we connect the last vertex of~$p$ with the first vertex of~$p'$.

\section{Main statement}    

    We call \emph{valuation} any (total) function~$\mu:\anyvarset\rightarrow\set{0,1}$ such that for every~$x\in X$, $\mu(\negvar{x})=1-\mu(x)$.
    The purpose
    of the remainder of this section is to show the following proposition,
    of which \rtheorem{walk-memb-untr} is a direct consequence.

    \begin{proposition}\lproposition{reduc-auto}
        There are exactly $N$ simple runs~$r$ in~$D_I\times \Ac$ such that~$\projD(r)=p_I$,
        where~$N$ is the number of distinct valuations that make~$I$ true.
    \end{proposition}
    
\section{Setting a valuation of the variables}

    Let us show that there is a bijection between the valuations and the runs in~$D_I\times \Ac$ for the part~$p_\text{setval}$.
    
    \begin{lemma}
        \begin{subthm}
            \item \llemma{reduc-auto-r-mu-i}
            For every valuation~$\mu$ there exists a simple run~$r_\mu$ in~$D_I\times \Ac$ such that: for every~$x\in X$, the run $r_\mu$ passes through the vertex~$(x,\mu(x))$.
             \item \llemma{reduc-auto-r-mu-ii}
             For every valuation~$\mu$, the run~$r_\mu$ passes through the vertices $((\anyvar{x},C_i),\mu(\anyvar{x}))$, for each~$x\in \anyvarset$ and $1\le i\le \gamma$.
            \item \llemma{reduc-auto-r-mu-iii}
            Let~$r$ be any run in~$D_I\times \Ac$ such that~$\projD(r)=p_\text{setval}$, 
        then~$r=r_\mu$ for some valuation~$\mu$.

        \end{subthm}
    \end{lemma}
    \begin{proof} \enumstyle{a}
        For each~$x\in X$, we denote by~$r_{\mu,x}$ the following run of~$D_I\times\Ac$.
        \begin{equation*}
            \begin{array}{@{}r@{}l@{}l@{}}
            r_{\mu,x} = (x^\text{in},\top) \rarrow{\varL} (x,\mu(x)) &{}\rarrow{\keepL}((x,C_0),\mu(x))\\
            &{}\rarrow{\keepL}\cdots\\
            &{}\rarrow{\keepL}((x,C_\gamma),\mu(x))\\
            {}\rarrow{\invertL}(\negvar{x},\mu(\negvar{x}))  & {}\rarrow{\keepL} ((\negvar{x},C_\gamma),\mu(\negvar{x}))\\
                   & {}\rarrow{\keepL}\cdots\\
                   & {}\rarrow{\keepL}((\negvar{x},C_0),\mu(\negvar{x}))\rarrow{\resetL}(x^\text{out},\top)
            \end{array}
        \end{equation*}
        Then we define~$r_\mu$ as follows.
        \begin{equation*}
            (\text{Start},\top)\rarrow{\resetL} r_{\mu,{x_1}} \rarrow{\resetL} \cdots \rarrow{\resetL} r_{\mu,{x_n}} \rarrow{\resetL} (\text{Mid},\top)
        \end{equation*}
        By construction, $r_{\mu,x}$ is a simple run of~$D_I\times\Ac$.
        \medskip
        
        \enumstyle{b} Follows from the definition of~$r_\mu$ in item \enumstyle{a}.
        
        \medskip
        
        \enumstyle{c} 
        For each variable~$x \in X$, the vertex~$x$ appears exactly once in~$\pi_\text{setval}$ hence there is exactly one occurrence in~$r$ of a vertex of the form~$(x,s)$, for some~$s\in \set{0,1,\top}$. 
        Note moreover that the edge coming into~$x$ in $p_\text{setval}$ is labelled by~$\varL$ hence that~$s\in\set{0,1}$ from the definition of~$\Ac$ (cf.\@ \rfigure{reduc-auto-var}).
        We let~$\mu$ denote the valuation that maps each~$x\in X$ to the unique~$s$ such that~$(x,s)$.
        
        From the definition of~$\Ac$ (cf.\@ \rfigure{reduc-auto}), the letters  $\resetL$, $\keepL$ and $\invertL$ are \emph{deterministic}, in the sense that there is at most one transition going out from every state and labelled by one of these letters.
        It follows that~$r=r_\mu$ since both runs coincide on all states coming after an edge labelled by~$\varL$.
    \end{proof}

\section{Checking the clauses}
    
    \begin{lemma}
        For each valuation~$\mu$,
        \begin{subthm}
            \item \llemma{reduc-auto-clause-i}
            if~$\mu$ makes~$I$ true, then there exists a unique run~$r'_\mu$
             such that both~$\pi(r'_\mu)=p_\text{checksat}$, and~$r^{}_\mu\cdot r'_\mu$ is simple;
            \item \llemma{reduc-auto-clause-ii}
            if~$\mu$ makes~$I$ false, then there is no run~$r'$ such that~$\projD(r')=p_\text{checksat}$ and~$r_\mu\cdot r'$ is simple.
        \end{subthm}
    \end{lemma}
    
    We say that two runs~$r$ and $r'$ are \emph{mutually simple} if~$r$ and~$r'$ have no vertex in common.
        
    \begin{proof}[Proof of \enumstyle{a}]
        Let~$C_i=\anyvar{x}_k\vee\anyvar{x}_\ell\vee\anyvar{x}_m$ be a clause of~$I$.
        By hypothesis, one of the atom is made true by~$\mu$.
        Let~$r'_{\mu,C_{i}}$ be the run in~$D_I\times\Ac$ defined as follows, depending on which among the atoms~$\anyvar{x}_k$, $\anyvar{x}_\ell$ and $\anyvar{x}_m$ is
        made true by~$\mu$
        \begin{alignat*}{2}
        \intertext{If~$\mu(\anyvar{x}_k)=1$:}
            r'_{\mu,C_{i}}= (C_i^\text{in},\top) 
            &{}\rarrow{\varL}  ((\anyvar{x}_k,C_i),{}   &0&) \\ 
            &{}\rarrow{\evalL} ((\anyvar{x}_\ell,C_i),{}&\top&)\\
            &{}\rarrow{\evalL} ((\anyvar{x}_m,C_i),{}   &\top&) \rarrow{\checkL} (C_i^\text{out},\top)\\
        \intertext{If~$\mu(\anyvar{x}_k)=0$ and $\mu(\anyvar{x}_\ell)=1$:}
            r'_{\mu,C_{i}}= (C_i^\text{in},\top) 
            &{}\rarrow{\varL}  ((\anyvar{x}_k,C_i),{}   &1&) \\ 
            &{}\rarrow{\evalL} ((\anyvar{x}_\ell,C_i),{}&0&)\\
            &{}\rarrow{\evalL} ((\anyvar{x}_m,C_i),{}   &\top&) \rarrow{\checkL} (C_i^\text{out},\top)
        \intertext{If~$\mu(\anyvar{x}_k)=\mu(\anyvar{x}_\ell)=0$ and $\mu(\anyvar{x}_m)=1$:}
            r'_{\mu,C_{i}}= (C_i^\text{in},\top) 
            &{}\rarrow{\varL}  ((\anyvar{x}_k,C_i),{}   &1&) \\ 
            &{}\rarrow{\evalL} ((\anyvar{x}_\ell,C_i),{}&1&)\\
            &{}\rarrow{\evalL} ((\anyvar{x}_m,C_i),{}   &0&) \rarrow{\checkL} (C_i^\text{out},\top)
        \end{alignat*}
        One may check that in each case, the vertex~$((\anyvar{x}_k,C_i),s)$ in $r_{\mu,C_{i}}$ is such that~$s=1- \mu(\anyvar{x_k})$; 
        and that the same is true for~$x_\ell$ and~$x_m$.
        It follows that~$r_\mu$ and~$r'_{\mu,C_{i}}$ are mutually simple.

        By definitions, $r_{\mu,C_{i}}$ and $r_{\mu,C_{j}}$ are mutually simple
        if~$i\neq j$.
        Hence, the run~$r^{}_\mu r'_\mu$ is simple where~$r'_\mu$ is defined as follows.
        \begin{equation*}
        r'_\mu ={}
            (\text{Mid},\top)\rarrow{\resetL}     r_{\mu,C_1}\rarrow{\resetL}\cdots\rarrow{\resetL}r_{\mu,\gamma}\rarrow{\resetL} \text{End}
        \end{equation*}
        It may be verified that~$r'_\mu$ is unique.
        The only letters that are nondeterministic in $\Ac$ are~$\varL$ and~$\evalL$,
        and the choice is between state~$0$ and~$1$.
        More precisely, these letters always bring up a choice
        between between vertex~$((\anyvar{x},C_i),0)$ or~$((\anyvar{x},C_i),1)$ in~$D_I\times\Ac$
        for some~$i\in\set{1,\ldots,\gamma}$ and some variable~$\anyvar{x}\in X\cup \negvarset$.
        One of those two vertices necessarily appears in~$r_\mu$ (as $((\anyvar{x}_k,C_i),\mu(\anyvar{x}))$), hence only the other one may appear in~$r'_\mu$.

\end{proof}
\begin{proof}[Proof of \enumstyle{b}]
        For the sake of contradiction, let us assume that there exists a run~$r'$
        such that~$\projD(r')=\pi_\text{checksat}$ and such that~$r_\mu r'$ is simple.
        Since $\mu$ makes~$I$ false, it makes $C_i=\anyvar{x}_k\vee\anyvar{x}_\ell\vee\anyvar{x}_m$ false for some~$i$.
        Let~$r'_{C_i}$ be the subwalk of~$r'$ such that~$\projD(r'_{C_i})=p_{C_i}$; it may we written as follows. 
        \begin{alignat*}{2}
            r'_{C_i}= (C_i^\text{in},s^\text{in}) 
            &{}\rarrow{\varL}  ((\anyvar{x}_k,C_i),{}   &&s_k) \\ 
            &{}\rarrow{\evalL} ((\anyvar{x}_\ell,C_i),{}&&s_\ell)\\
            &{}\rarrow{\evalL} ((\anyvar{x}_m,C_i),{}   &&s_m) \rarrow{\checkL} (C_i^\text{out},s^\text{out})
        \end{alignat*}
        for some $s^\text{in},s_k,s_\ell, s_m,s^\text{out}\in\set{0,1,\top}$.
        Since~$\mu$ makes~$I$ false, it follows that~$\mu(\anyvar{x}_k)=0$, hence from \rlemma{reduc-auto-r-mu-ii} that~$r_\mu$ contains the vertex~$((\anyvar{x}_k,C_i),0)$.
        Moreover, since~$r_\mu r'$ is simple, it follows that  $s_k\neq 0$.
        A similar reasoning yields that~$s_\ell\neq 0$ and~$s_m\neq 0$
        Regardless of the value of~$s^\text{in}$, it follows from the transitions of~$\Ac$ for letter~$\varL$ (cf.\@ \rfigure{reduc-auto-var}) that~$s_k\in\set{0,1}$, hence~$s_k=1$; it then follows from the transitions of~$\Ac$ for letter~$\evalL$ (cf.\@ \rfigure{reduc-auto-eval}) that~$s_\ell=1$, and with the same argument that~$s_m=1$.
        Since~$\Ac$ has no transition labelled by $\checkL$ and going out from state~$1$ (cf.\ \rfigure{reduc-auto-check}), this leads to a contradiction.
    \end{proof}

\section{Proof of Proposition \ref{p.reduc-auto}}

Let~$N$ be the number of valuations that make~$I$ true.
Let~$\mu$ be a valuation that makes~$I$ true. \rlemma{reduc-auto-r-mu-i} yields a run $r_\mu$ and \rlemma{reduc-auto-clause-i} yields a run $r'_\mu$ such that $r_\mu \cdot r'_\mu$ is simple and ${\projD(r_\mu\cdot r'_\mu)=p_I}$.

Note that if~$\mu,\mu'$ denote two valuations that makes~$I$ true,~$r_\mu=r_\mu'$ implies that for every~$x\in X$, $\mu(x)=\mu'(x)$ (due to the condition in \rlemma{reduc-auto-r-mu-i}), hence that $\mu=\mu'$.
Thus, the previous paragraph defines~$N$ distinct simple runs in~$D_I\times\Ac$.

Let~$r\cdot r'$ be any simple run in~$D_I \times\Ac$ such that~$\projD(r)=p_\text{setval}$ and~$\projD(r')=p_\text{checksat}$. 
Hence from \rlemma{reduc-auto-r-mu-iii}, there exists~$\mu$ such that~$r=r_\mu$.
It is impossible that~$\mu$ makes~$I$ false since the existence of~$r'$ would be in contradiction with~\rlemma{reduc-auto-clause-ii}.
Hence,~$\mu$ makes~$I$ true and the unicity in~\rlemma{reduc-auto-clause-i} implies that~$r'=r'_\mu$.
\hspace*{0cm plus 1fill}\qedsymbol

\begin{remark}
The size of the database~$D_I$ built in \rsection{db-B} may be up to quadratic in the size of the 3-SAT instance~$I$, namely in $O(\gamma n)$.
This is due to the fact that we create a lot of "useless" vertices. In particular, every vertex~$(\anyvar{x},C_i)$ can be omitted if $\anyvar{x}$ does not appear in clause $C_i$.
By omitting those vertices, the database~$D_I$ would be of a size in $O(\gamma+n)$, at the cost of making the definition more involved.
\rfigure{reduc-graph}, \pfigure{reduc-graph}, gives an example of that simplified~$D$.
\end{remark}

%% file: figure/reduc-auto-var.tex
\begin{tikzpicture}
\input{figure/reduc-auto-base}
\path[edge] (0) to[bend left] node[above] {\varL} (1);
\path[edge] (1) to[bend left] node[below] {\varL} (0);
\path[edge] (top) to node[above,sloped] {\varL} (1);
\path[edge] (top) to node[above,sloped] {\varL} (0);

\path[south loop] (0) to node[below] {\varL} (0);
\path[south loop] (1) to node[below] {\varL} (1);

\end{tikzpicture}

%% file: figure/reduc-auto-keep.tex
\begin{tikzpicture}
\input{figure/reduc-auto-base}

\path[south loop] (0) to node[below] {\keepL} (0);
\path[south loop] (1) to node[below] {\keepL} (1);
\path[north loop] (top) to node[above] {\keepL} (top);

\end{tikzpicture}

%% file: figure/reduc-auto-invert.tex
\begin{tikzpicture}
\input{figure/reduc-auto-base}
\path[edge] (0) to[bend left] node[above] {\invertL} (1);
\path[edge] (1) to[bend left] node[below] {\invertL} (0);
\end{tikzpicture}

%% file: figure/reduc-auto-eval.tex
\begin{tikzpicture}
\input{figure/reduc-auto-base}
\path[edge] (1) to node[below] {\evalL} (0);

\path[edge] (0) to 
node[sloped,above] {\evalL} (top);

\path[south loop] (1) to node[below] {\evalL} (1);
\path[north loop] (top) to node[above] {\evalL} (top);

\end{tikzpicture}

%% file: figure/reduc-auto-check.tex
\begin{tikzpicture}
\input{figure/reduc-auto-base}

\path[edge] (0) to  node[sloped,above] {\checkL} (top);

\path[north loop] (top) to node[above] {\checkL} (top);

\end{tikzpicture}

%% file: figure/reduc-auto-reset.tex
\begin{tikzpicture}
\input{figure/reduc-auto-base}
\path[edge] (1) to node[above,sloped] {\resetL} (top);
\path[edge] (0) to node[above,sloped] {\resetL} (top);

\path[north loop] (top) to node[above] {\resetL} (top);

\end{tikzpicture}

%% file: appendix/topo_coding.tex
\newappendix{Glushkov construction provides hard instance for \problemfont{Walk Membership}}
\label{a.topo_coding}

The purpose of appendix~\theappendix{} is to prove the proof of \rtheorem{glus-does-not-help}, below,
which is the center piece behind \rproposition{bindingtrail-simplerun-equivalence}.
The proof uses the novel notion of \emph{topological coding} of an automaton, which we plan to flesh-out in a future independent document; we leave a preview of it here in order for this preprint to be self-contained.

\begin{theorem}\ltheorem{glus-does-not-help}
    There exists a fixed expression~$R$ such that \problemfont{Walk Membership} is NP-hard for~$\Ac=\gl(R)$ under simple-run semantics.
\end{theorem}

\rsection{glus-aut} introduces a different, equivalent definition for the Glushkov automaton of an expression. It allows for more effective notations, whereas the initial definition was only stated in a declarative way. 
\rsection{topo-codi} defines \emph{topological codings} and \rsection{topo-codi-prop} gives a precise meaning to the intuition that a topological coding somehow simulates another automaton. 
\rsection{topo-codi-theo} states and shows \rtheorem{topo-codi-glus}, the main result of appendix~\theappendix: any automaton can be encoded into a Glushkov automaton.
\rsection{glus-does-not-help-proo} applies \rtheorem{topo-codi-glus} to the proof of \rtheorem{glus-does-not-help}.
Finally \rsection{proof-prop} briefly explains how a counterpart to \rtheorem{glus-does-not-help} can be proved under binding-trail semantics.

\section{Glushkov automaton}
\lsection{glus-aut}


\newcommand{\annot}[2]{\big[\smash{\begin{array}{@{}c@{}}\scriptstyle#1\\[-1.2ex]\scriptstyle#2\end{array}}\big]}

A linearisation of an expression~$R$ over an alphabet $\Gamma$ is a pair~$\langle \Gamma,R'\rangle$ where
\begin{itemize}
    \item $\Gamma$ is a finite set of \emph{annotations};
    \item $R'$ is an expression over~$\Sigma\times\Gamma$ such that
    \begin{itemize}
        \item every letter in~$\Sigma\times\Gamma$ appears at most once in~$R'$
        \item $f(R')=R$ where $f$ is the projection $\Sigma\times\Gamma\rightarrow\Sigma$ on the first component, lifted to regular expressions.
    \end{itemize}
\end{itemize}
We denote the letter $(a,i)\in\Sigma\times\Gamma$ as $\annot{a}{i}$.
Classically, one linearises~$R$ using~$\Gamma=\{1,\ldots,n\}$, where~$n$ is the number of atoms in~$R$, annotating the $i$-th leftmost atom in~$R$ with~$i$.
For instance, the linearisation of~$b^*(ab^*ab^*)^*$ would be $\langle R',\Gamma\rangle$ with:
$$
    R'={} \annot{b}{1}^*\left(\annot{a}{2}\annot{b}{3}^*\annot{a}{4}\annot{b}{5}^*\right)^* \qquad \text{ and }\qquad
    \Gamma={} \set{1,2,3,4,5}.
$$

Given a regular expression~$R$ over an alphabet~$\Sigma$, the \emph{Glushkov automaton} associated with~$R$, denoted by $\gl(R)\stdaut{}$, is defined as follows from any linearisation $\langle\Gamma,R'\rangle$ of~$R$.
\begin{align}
    Q={}& \set{\mathsf{init}}\cup \Sigma\times\Gamma 
    \label{eq:glus-stat}
    \\[5pt]
    \Delta={}& \setst*{\left(\annot{a}{i},b,\annot{b}{k}\right)}{\begin{array}{@{}l@{}}
    \annot{a}{i},\annot{b}{k}\in \Sigma\times\Gamma\\
    \exists w,w' \text{ such that }w\annot{a}{i}\annot{b}{k}w'\in \lang(R'))
    \end{array}
    }
    \notag
    \\
    &~\bigcup \setst*{(\mathsf{init},a,\annot{a}{i})}%
        {\begin{array}{@{}l@{}}
        \annot{a}{i}\in\Sigma\times\Gamma\\
        \exists w \text{ such that }\annot{a}{i}w\in \lang(R')\\
        \end{array}}
    \label{eq:glus-tran}
    \\[5pt]
    I={}& \set{\mathsf{init}}
    \label{eq:glus-init}
    \\
    F={}& \setst*{\annot{a}{i}}%
        {\exists w \text{ such that }w\annot{a}{i}\in \lang(R')}
    \label{eq:glus-fina}
\end{align}

\section{Topological coding of an automaton}
\lsection{topo-codi}

The notion of topological coding is an adaptation to automata of the notion of \emph{topological minor} for directed graphs  \cite{DBLP:books/daglib/0030488}.
%
Topological codings are defined formally in \rdefinition{topo-codi}. We give a first intuitive definition below.


Intuitively, $\Bc$ is a topological coding of $\Ac$ if $\Bc$ may be built from~$\Ac$ by the following process:
\begin{itemize}
    \item for each letter~$a$ in~$\alphabetof{\Ac}$, the alphabet of~$\Ac$, choose a nonempty word~$\lambda(a)$ over~$\alphabetof{\Bc}$, the alphabet of~$\Bc$;
    \item replace each transition $(s,q,t)$ in~$\Ac$ by a fresh walk
    labelled by~$\lambda(a)$ that starts at $s$ and ends at $t$;
    \item optionally, choose a word~$u_i$ over~$\alphabetof{\Bc}$, add a fresh initial state~$init$, and for each initial state~$s$ of~$\Ac$, remove its initial status and add a fresh walk going from~$init$ to~$s$ 
    \item optionally, proceed similarly for final states with a word~$u_f$ and a fresh final state~$final$;
    \item then, one may add states and transitions, as long as it does not create a walk labelled by a word in $\im(\lambda)$, or a walk starting from an initial (resp.\@ ending in a final) state labelled by $u_i$ (resp.~$u_f$).
\end{itemize}

Given an automaton~$\Ac$, we let $\computof\Ac$ denote the set of computations in~$\Ac$.

\begin{definition}\ldefinition{topo-codi}
    We say that~$\stdaut{\Bc}$ is a \emph{topological out-coding} of~$\stdaut{\Ac}$, or simply a \emph{topological coding}\footnote{The \emph{out-} comes from the fact that the definitions of the~$W^\text{something}_\Bc$'s are not symmetric: $W_\Bc$ contains the walks going \textbf{out} of a state...} of~$\Ac$,
    if there exist:
     \begin{subthm}
        \item \ldefinition{topo-codi-star} 
        \ldefinition{topo-codi-words}
        two words~$u_i,u_f\in\alphabetof{\Bc}^*$;
        
        \item \ldefinition{topo-codi-lamb}
        an injective function~$\lambda:\alphabetof{\Ac}\rightarrow\alphabetof{\Bc}^+$
        
        \item \ldefinition{topo-codi-nu}
        an injective function~$\nu:\statesof{\Ac}\rightarrow \statesof{\Bc}$.

    \end{subthm}
    and, denoting
    \begin{align}
        W_\Bc^{\text{initial}} ={}
        & \setst*{w\in\computof{\Bc}}{\begin{array}{@{}l@{}}
        \src(w)\in\initialsof\Bc\\
        \lbl(w)=u_i\end{array}}
        \label{eq:w-b-init}
        \\
        W_\Bc^{\text{transition}} ={}
        & \setst*{w\in\computof\Bc}{\begin{array}{@{}l@{}}\src(w)\in\im(\nu)\\\lbl(w)\in\im(\lambda)\end{array}}
        \label{eq:w-b-tran}
        \\
        W_\Bc^{\text{final}} ={}
        & \setst*{w\in\computof\Bc}{\begin{array}{@{}l@{}}\src(w)\in\im(\nu)\\
        \lbl(w)=u_f\end{array}}
        \label{eq:w-b-fina}
    \end{align}
    there exist:
    \begin{subthm}[resume]
        
        \item \ldefinition{topo-codi-init}
        a bijection~$\finit:\initialsof{\Ac}\rightarrow W_\Bc^{\text{initial}}$
        such that 
        \begin{itemize}
            \item for every~$s\in\initialsof{\Ac}$, the walk $\finit(s)$ 
        ends in~$\nu(s)$
            \item for every~$w\in\im(\finit)$ and every state~$s$ that appears in~$w$ at a position that is not the last one, then~$s\notin \im(\nu)$;
        \end{itemize}
            
        \item \ldefinition{topo-codi-tran}
        a bijection~$\eta:\transitionsof{\Ac}\rightarrow W_\Bc^{\text{transition}}$
        such that for every~$(s,a,t)$ in $\transitionsof{\Ac}$, 
        the walk~$w=\eta((s,a,t))$ satisfies:
        \begin{itemize}
            \item $\eta((s,a,t))$ starts in~$\nu(s)$
            \item $\eta((s,a,t))$ is labelled by~$\lambda(a)$
            \item $\eta((s,a,t))$ ends in~$\nu(t)$
            \item every internal state\footnote{A state in a computation is \emph{internal} if it appears at a position that is not the first or the last one.} in~$\eta((s,a,t))$ is not in $\im(\nu)$;
        \end{itemize}

        \item \ldefinition{topo-codi-fina}
        a bijection~$\ffinal:\finalsof{\Ac}\rightarrow W_\Bc^{\text{final}}$ such that
        \begin{itemize}
            \item for every~$s\in\finalsof\Ac$, the walk $\ffinal(s)$ ends in~$F_\Bc$
            \item for every~$w\in\im(\ffinal)$ and every state~$s$ that appears in~$w$ at a position that is not the first one, then~$s\notin \im(\nu)$.
        \end{itemize}
        
    \end{subthm}
%
%
%
that collectively satisfy:
    \begin{subthm}[resume]
        \item \ldefinition{topo-codi-inje}
        for every~$w_1,w_2\in\big(\im(\finit)\cup\im(\eta)\cup\im(\ffinal)\big)$, if $w_1$ and $w_2$ have a transition or an internal state in common, then $w_1=w_2$.

        \ldefinition{topo-codi-end}
    \end{subthm}
\end{definition}

\section{Properties of a topological coding}
\lsection{topo-codi-prop}

In this section, we fix two automata~$\stdaut{\Ac}$ and~$\stdaut{\Bc}$ such that $\Bc$ is a \emph{topological out-coding} of~$\Ac$ and we reuse the notations of \rdefinition{topo-codi}.

\begin{remark}The properties below follow from the definition of a topological coding.
    \begin{subthm}
        \item The conditions \rdefinition*{topo-codi-init} and \rdefinition*{topo-codi-tran} imply that~$W_\Bc^{\text{transition}}$ and~$W_\Bc^{\text{initial}}$ are disjoint.
        Indeed, if~$u_i\neq \varepsilon$, the first state in each walk in $W_\Bc^{\text{transition}}$ is in~$\im(\nu)$, while the first state in each walk in~$W_\Bc^{\text{initial}}$ is not; and if~$u_i=\varepsilon$, all walks in~$W_\Bc^{\text{initial}}$ has length~$0$ while all walks in~$W_\Bc^{\text{transition}}$ have a positive length since $\len(\lambda(a))>0$ for every~$a\in \alphabetof{\Ac}$.
        
        \item Similarly,~$W_\Bc^{\text{transition}}$ and~$W_\Bc^{\text{final}}$ are disjoint. 
        
        \item Similarly,~$W_\Bc^{\text{initial}}$ and~$W_\Bc^{\text{final}}$ are disjoint unless~$u_i=u_f=\varepsilon$ and~$\initialsof\Ac\cap\finalsof\Ac\neq\emptyset$.

    \end{subthm}
\end{remark}

As expected, there is a strong correspondence between computations in an automaton and its topological coding.

\begin{lemma}\llemma{topo-codi-comp-bije}
    Let~$\Ac$ and~$\Bc$ be two automata such that~$\Bc$ is a topological coding of~$\Ac$.
    \begin{subthm}
    \item \llemma{topo-codi-comp-bije->}
    Let~$s_0 \xrightarrow{a_1} s_1 \xrightarrow{a_2} \cdots \xrightarrow{a_k} s_k$ be a computation in~$\mathcal{A}$.
    Then, $\nu(s_0)\xrightarrow{\lambda(a_1)}\nu(s_1)\xrightarrow{\lambda(a_2)}\cdots\xrightarrow{\lambda(a_k)}\nu(s_k)$ is a computation in~$\mathcal{B}$.
    
    \item \llemma{topo-codi-comp-bije<-}
    Conversely, let $\pi_{\mathcal{B}}$ be a computation in~$\mathcal{B}$ of the form \begin{equation*}
        \pi_{\mathcal{B}}=\nu(s_0)\xrightarrow{\lambda(a_1)}\nu(s_1)\xrightarrow{\lambda(a_2)}\cdots\xrightarrow{\lambda(a_k)}\nu(s_k)
    \end{equation*}
    then~$s_0\xrightarrow{a_1}s_1\xrightarrow{a_2}\cdots\xrightarrow{a_k}s_k$ is a computation in~$\mathcal{A}$.
    \end{subthm}
\end{lemma}
\begin{proof}
    Item~\enumstyle{a} follows from condition~\rdefinition*{topo-codi-tran} applied to each transition of~$\transitionsof{\Ac}$.
    Item~\enumstyle{b} follows from the fact that each walk
    $\nu(s_i)\rarrow{\lambda(a_{i+1})}\nu(s_{i+1})$ belongs to $W_{\Bc}^{\text{transition}}$, which allows to apply~\rdefinition*{topo-codi-tran}
    and concludes the proof.
\end{proof}

\begin{definition}\ldefinition{corr-comp}
Let~$\Bc$ be a topological coding of~$\Ac$. Given a successful
computation $\pi_\Ac=s_0 \xrightarrow{a_1} s_1 \xrightarrow{a_2} \cdots \xrightarrow{a_k} s_k$  in~$\Ac$,
we call \emph{corresponding computation} in~$\Bc$, the computation $\pi_\Bc$ defined as
\begin{equation*}\newcommand{\myrule}{\rule[-2.5ex]{0pt}{6ex}}
        \overbrace{\myrule x\rarrow{u_i}\nu(s_0)}^{\finit(s_0)}
        \setlength{\dimen0}{0pt-\widthof{$\nu(s_0)$}}\hspace{\dimen0}
        \underbrace{\myrule
        \setlength{\dimen0}{\widthof{$\nu(s_0)$}}\hspace*{\dimen0}\rarrow{\lambda(a_1)}\nu(s_1)}_{\eta((s_0,a_1,s_1))\mathrlap{\hspace*{17mm}\displaystyle\cdots}}\rarrow{\lambda(a_2)}~\cdots~ 
        \rarrow{\lambda(a_{k-1})}
        \underbrace{\myrule\rule[-1.8ex]{0pt}{0pt}\nu(s_{k-1}) \rarrow{\lambda(a_k)} \nu(s_k)}_{\eta((s_{k-1},a_1,s_k))}
        \setlength{\dimen0}{0pt-\widthof{$\nu(s_k)$}}\hspace{\dimen0}\overbrace{\myrule\setlength{\dimen0}{\widthof{$\nu(s_k)$}}\hspace*{\dimen0}\rarrow{u_f} y}^{\ffinal(s_k)}
\end{equation*}
where~$x=\src(\finit(s_0))$ and~$y=\tgt(\ffinal(s_k))$.
\end{definition}

We show that there is a bijection between successful computations in $\Ac$ and successful computations in $\Bc$ of a specific shape.

\begin{proposition}
Let~$\Bc$ be a topological coding of~$\Ac$.
    \begin{subthm}
        \item \llemma{corr-run-succ}
        Let~$\pi_\Ac$ be a successful computation in~$\Ac$, then the corresponding computation $\pi_\Bc$ in~$\Bc$ is successful.
        \item \llemma{inv-run-succ}        Let~$\pi_\Bc$ be a successful computation in~$\Bc$ such that~$\lbl(\pi_\Bc)=u_i\,\lambda(u)\,u_f$ for some~$u\in\alphabetof{\Ac}^*$.
        Then there exists a successful computation in~$\Ac$, of which~$\pi_\Bc$ is the corresponding computation.
        \item     For every word~$u\in\alphabetof{\mathcal{A}}^*$,
            $u$ is accepted by~$\mathcal{A}$ if and only if~$u_i\,\lambda(u)\,u_f$ is accepted by~$\mathcal{B}$.
    \end{subthm}
\end{proposition}
\begin{proof}
    Item \enumstyle{a} follows from \rdefinition{topo-codi-init} which ensures that $\src(\pi_\Bc)\in\initialsof{\Bc}$ and \rdefinition{topo-codi-fina} which ensures that $\tgt(\pi_\Bc)\in\finalsof{\Bc}$.
    
    Item \enumstyle{b}. 
    We write~$u=a_1\cdots a_k$.
    The computation~$\pi_\Bc$ may be factorised as ~$\pi_\Bc=\pi_0\pi_1\cdots\pi_k\pi_{k+1}$
    with $\lbl(\pi_0)=u_i$, $\lbl(\pi_{k+1})=u_f$, and
    for every $i$, $0<i\leq k$, $\lbl(\pi_i)=\lambda(a_i)$.
    Since~$\pi_\Bc$ is successful, $\pi_0$ is such that  
    \begin{equation*}
        \src(\pi_{0})\in\initialsof{\Bc} \quad\text{and}\quad \lbl(\pi_{0})=u_i
    \end{equation*}
    which implies that~$\pi_0\in W_\Bc^{\text{initial}}$ by definition, and $\tgt(\pi_0)\in\im(\nu)$ from the first item of Condition \rdefinition*{topo-codi-init}.
    Then a simple induction using the third item of Condition \rdefinition*{topo-codi-tran} yields that for every $i$, $0<i\leq k$,  $\pi_i\in W_\Bc^{\text{transition}}$ and~$\tgt(\pi_i)\in\im(\nu)$.
    Finally since~$\pi_{k+1}$ is such that 
    \begin{equation*}
        \src(\pi_{k+1})\in\im(\nu) \quad\text{and}\quad \lbl(\pi_{k+1})=u_f
    \end{equation*}
    it holds~$\pi_{k+1}\in W_\Bc^{\text{final}}$. We let~$\pi_{\Ac}$ denote the following computation.
    \begin{equation*}
        \pi_\Ac=\finit^{-1}(\pi_0) \cdot \eta^{-1}(\pi_1)\cdots \eta^{-1}(\pi_k) \cdot \ffinal^{-1}(\pi_{k+1})
    \end{equation*}
    Conditions \rdefinition*{topo-codi-init}, \rdefinition*{topo-codi-tran} and \rdefinition*{topo-codi-fina} ensure that~$\pi_\Ac$ is a well-defined computation in $\Ac$ and that it is successful.

    Item \enumstyle{c} follows directly from \enumstyle{a} and \enumstyle{b}.
\end{proof}

\section{Main statement}
\lsection{topo-codi-theo}

\begin{theorem}\ltheorem{topo-codi-glus}
    Let~$\Ac\stdaut{}$ be a trim automaton.
    There exists a regular expression $R$ such that $\gl(R)$ is a topological coding of $\Ac$.

    Moreover, let $m=\card(\Delta)$. The size of $R$ is in $O(m^2)$, there is exactly one Kleene star in $R$, and the number of transitions in~$\gl(R)$ is in $O(m^2)$.
\end{theorem}

The remainder of section \thesection{} is dedicated to the proof of \rtheorem{topo-codi-glus}.


    We write~$\stdaut{\Ac}$ and~$m=\card(\transitionsof{\Ac})$.
    We let~$G$ denote any bijection $G:\transitionsof\Ac\rightarrow\set{1,\ldots,m}$.
    We let $H$ denote the only bijection $H:\transitionsof\Ac\rightarrow\set{1,\ldots,m}$ that meets the following.
    \begin{equation}
        \forall e\in\transitionsof\Ac,\quad G(e)+H(e)=m+1
    \end{equation}
    
    
    For each state~$q\in Q$ we let $R^\text{left}_q$ and
    $R^\text{right}_q$ denote the following expressions.
    \begin{align}
        R^\text{left}_q ={}& \overbrace{\rule{0pt}{1.5ex}\varepsilon~~~~+}^{\text{~if $q\in \initialsof{\Ac}$}}\sum_{\substack{s\in Q,~a\in\varSigma\\ e=(s,a,q)\in E}} \overbrace{\rule{0pt}{1.5ex}a\cdots a}^{G(e)\text{ times}}\\[10pt]
        R^\text{right}_q ={}&
                \overbrace{\rule{0pt}{1.5ex}\varepsilon~~+}^{\text{~if $q\in \finalsof{\Ac}$}}\sum_{\substack{a\in\varSigma,~t\in Q\\ e=(q,a,t)\in E}} \overbrace{\rule{0pt}{1.5ex}a\cdots a}^{H(e)\text{ times}}
    \end{align}
    Finally, given a fresh letter  $\sigma\notin\alphabetof\Ac$,
    we define the expression~$R$ over $\alphabetof\Ac\uplus\set{\sigma}$ as follows:
    \begin{equation}
        R=
        \left(
        \sum_{q\in Q}R^{\text{left}}_q \cdot \sigma \cdot R^{\text{right}}_q
        \right)^*
    \end{equation}
    \noindent Note that for $R$ to be well defined, we need the automaton~$\Ac$ to be trim.  Indeed, an automaton that is not accessible might feature a state~$q$ that is not initial and that has no incoming transition. In that case, the subexpression $R^\text{left}_q$ would be an empty sum, and the neutral element for the sum of regular expression is not allowed in our formalism for regular expressions.
    A similar phenomenon occurs would~$\Ac$ not be coaccessible.
    
    \medskip Now, let us show that $\gl(R)$ is a topological coding of $\Ac$.
    We define a particular linearisation $\bar{R}$ of $R$ in order to be able to explicitly state the elements $u_i$, $u_f$, $\lambda$, $\nu$, $\finit$, $\eta$ and $\ffinal$
    that realise the the topological coding.
    
    \medskip
    Let~$\langle \bar{R},\Gamma\rangle$ be  the linearisation of~$R$ with $\bar{R}$ defined as follows and $\Gamma$ defined implicitly .
       \begin{align}
        \bar{R}={}&
        \left(
        \sum_{q\in Q}\bar{R}^{\text{left}}_q \cdot \annot{\sigma}{q} \cdot \bar{R}^{\text{right}}_q
        \right)^*
        \label{eq:R-bar}
        \\
    \intertext{where:}
        \bar{R}^\text{right}_q ={}&
         \overbrace{\rule{0pt}{1.5ex}\varepsilon~~+}^{\text{~if $q\in \finalsof{\Ac}$}}\sum_{\substack{a\in\varSigma,~t\in Q\\ e=(q,a,t)\in E}} \annot{a}{e,0}\cdots \annot{a}{e,H(e)-1}
         \label{eq:R-bar-right}
         \\[10pt]
        \bar{R}^\text{left}_q ={}& \overbrace{\rule{0pt}{1.5ex}\varepsilon~~~~+}^{\text{~if $q\in \initialsof{\Ac}$}}\sum_{\substack{s\in Q,~a\in\varSigma\\ e=(s,a,q)\in E}} \annot{a}{e,H(e)}\cdots \annot{a}{e,m}
        \label{eq:R-bar-left}
    \end{align}
    
    Notice that, in $\bar{R}^\text{left}_q$ \eqref{eq:R-bar-left}, there are indeed $G(e)$ concatenated atom in the member of the sum corresponding to transition~$e$ since~$G(e)+H(e)=m+1$.
 
    \medskip
    In the following, $\stdaut{\Bc}$ denotes the Glushov automaton built from the linearisation~$\langle \bar{R},\Gamma\rangle$.      
    Intuitively, each state $q$ of $\Ac$ is encoded by the state $\annot{\sigma}{q}$ of $\Bc$.
    
    \medskip 
    Using notations from \rdefinition{topo-codi}, we now prove that~$\Bc$ is a topological coding of~$\Ac$. We define the words~$u_i,u_f$ as $u_i = \sigma$ and $u_f=\varepsilon$ and the function~$\lambda,\nu,\finit,\eta,\ffinal$ as follows.
    \begin{align}
        \lambda:&& \alphabetof\Ac \quad\rightarrow\quad& \alphabetof\Ac\uplus\set{\sigma}   
        \notag\\
                && a \quad\mapsto\quad& a^{m+1}\sigma
        \lequation{lambda}
        \displaybreak[1]\\[10pt]
        \nu :&& \statesof{\Ac} \quad\rightarrow\quad& \statesof{\Bc}
        \notag\\
             && q \quad\mapsto\quad& \annot{\sigma}{q}
        \lequation{nu}
        \displaybreak[1]\\[10pt]
        \finit : && \initialsof{\Ac} \quad\rightarrow\quad& \comput(\Bc)
        \notag\\
            && q \quad\mapsto\quad& \mathsf{init}\rarrow{\sigma}\annot{q}{\sigma}
        \lequation{i}
        \displaybreak[1]\\[10pt]
        \eta:&& \transitionsof{\Ac}\quad\rightarrow\quad& \comput(\Bc)
        \notag\\ 
             && e\quad\mapsto\quad&\annot{\sigma}{q} \rarrow{a} \annot{a}{e,0}\rarrow{a}
             \annot{a}{e,1}
             \cdots\rarrow{a}\annot{a}{e,m}\rarrow{\sigma} \annot{\sigma}{q'}
        \lequation{eta}
        \intertext{where~$e=(q,a,q')$}
        \ffinal:  && \finalsof{\Ac} \quad\rightarrow\quad& \comput(\Bc)
        \notag\\
            && q \quad\mapsto\quad& \annot{\sigma}{q}
        \lequation{f}
    \end{align}
    Notice that the image of~$\ffinal$ are computations of length~$0$, hence reduced to a single state.
    
    It remains to show that conditions of~\rdefinition{topo-codi} are satisfied.
    
     \paragraph*{Conditions \protect\rdefinition*{topo-codi-words} to \protect\rdefinition*{topo-codi-nu}}
    Condition \rdefinition*{topo-codi-words} is trivially satisfied,
    and it is easy to see that~$\lambda$ and~$\nu$ are injective;
    hence Conditions \rdefinition*{topo-codi-lamb} and \rdefinition*{topo-codi-nu} hold.
    
    
    \paragraph*{Condition \protect\rdefinition*{topo-codi-init}}
  
    Since~$\initialsof{\Bc}=\set{\mathsf{init}}$ and $u_i=\sigma$,
    it follows from~\eqref{eq:w-b-init} and~\eqref{eq:glus-tran}:
    \begin{align*}
        W_{\Bc}^\text{initial} = {}& \setst*{w\in\computof{\Bc}}{\src(w)\in \initialsof\Bc \text{ and } \lbl(w)=u_i}
        \\
        ={}& \setst*{\mathsf{init}\rarrow{\sigma}\annot{\sigma}{q}}{
            \begin{array}{@{}l@{}}
                q\in \statesof{\Ac}\\
                \text{$\varepsilon\in L(\bar{R}_q^\text{left})$}
            \end{array}}
        \\
        ={}& \setst*{\mathsf{init}\rarrow{\sigma}\annot{\sigma}{q}}{q\in \initialsof{\Ac}}
        \\
        ={}& \im(\finit)
    \end{align*}
    Hence~$\finit$ is a bijection~$\initialsof{\Ac}\rightarrow W_{\Bc}^\text{initial}$. 
    Then, it may be verified that Condition \rdefinition*{topo-codi-init} is met by definition.
    
      
    
    \paragraph*{Condition \protect\rdefinition*{topo-codi-tran}}
    Showing this condition  amounts to showing the following.
    \begin{equation}\label{eq:goal-tran}
        W_\Bc^{\text{transition}}  = \im(\eta) 
    \end{equation}        
    Indeed, all other requirements follow from the definition of~$\eta$.
    It is clear that~$\im(\eta)\subseteq W_\Bc^{\text{transition}}$ so let us show the other direction.
    Let~$w\in W_\Bc^{\text{transition}}$, hence 
    \begin{itemize}
        \item there exists~$q_1\in\statesof{\Ac}$ such that~$\src(w)=\annot{\sigma}{q_1}$ 
        \item there exists~$a\in\alphabetof\Ac$ such that~$\lbl(w)=a^{m+1}\sigma$ 
        \item since~$\lbl(w)$ ends with the letter~$\sigma$, there exists~$q_2\in\statesof{\Ac}$ such that~$\tgt(w)=\annot{\sigma}{q_2}$. 
    \end{itemize}
    Thus, the computation~$w$ is of the form $w = w_1w_2w_3$ with $\lbl(w_1)\in L(\bar{R}^\text{right}_{q_1})$, $\lbl(w_2)\in L(\bar{R}^\text{left}_{q_2})$ and $\lbl(w_3) = \sigma$. Let $\ell_1,\ell_2\in\mathbb{N}$ such that $\lbl(w_1) = a^{\ell_1}$ and $\lbl(w_2) = a^{\ell_2}$.
    
    If $\ell_1 > 0$ and $\ell_2 > 0$, there exist~$e_1=(q_1,a,t)\in\transitionsof{\Ac}$
    and $e_2=(s,a,q_2)\in\transitionsof{\Ac}$ that satisfy:
    \begin{multline*}
        w = \annot{\sigma}{q} 
        \rarrow{}\annot{a}{e_1,0} 
        \rarrow{a} \annot{a}{e_1,1} 
        \rarrow{a} \cdots 
        \rarrow{a} \annot{a}{e_1,H(e_1)-1} 
        \\
        \rarrow{a} \annot{a}{e_2,H(e_2)}
        \rarrow{a} \annot{a}{e_2,H(e_2)+1}
        \rarrow{a} \cdots
        \rarrow{a} \annot{a}{e_2,m} \rarrow{\sigma} \annot{\sigma}{q'}
    \end{multline*}
    The only way for the label of~$w$ to be~$a^{m+1}\sigma$ 
    is if~$H(e_1)=H(e_2)$, that is if~$e_1=e_2$.
    It follows that~$(q_1,a,q_2)\in\transitionsof{\Ac}$ and one may verify that
    $w=\eta((q_1,a,q_2))$.
    
    Otherwise, if $\ell_1 = 0$, then $\ell_2 = m+1$. This is a contradiction with $\lbl(w_2)\in L(\bar{R}^\text{left}_{q_2})$ because all words in $L(\bar{R}^\text{left}_{q_2})$ are of length at most $m$. The case where $\ell_2 = 0$ is impossible for similar reasons.
    
    
    
    \paragraph*{Condition \protect\rdefinition*{topo-codi-fina}}
    Since~$u_f=\varepsilon$, Condition \rdefinition*{topo-codi-fina} amounts to showing 
    that~$\setst*{\annot{\sigma}{q}}{q\in\finalsof{\Ac}}\subseteq \finalsof{\Bc}$.
    It is true from the definition of~$\bar{R}$: indeed $\varepsilon\in L(\bar{R}^\text{right}_q)$ if and only if~$q\in\finalsof{\Ac}$.
    
    
    
    \paragraph*{Condition \protect\rdefinition*{topo-codi-inje}}
    Let~$w_1,w_2\in\big(\im(\finit)\cup\im(\eta)\cup\im(\ffinal)\big)$ such that
    $w_1$ and $w_2$ have a transition or an internal state in common.
    The walks in~$\im(\ffinal)$ have no transitions nor internal states, 
    hence~$w_1,w_2\in\big(\im(\finit)\cup\im(\eta))$.
    Since the walks in~$\im(\finit)$ consists of a single transition and that transition
    is never used by any walk in~$\im(\eta)$, then either~$w_1,w_2\in\im(\finit)$ or~$w_1,w_2\in\im(\eta)$.
    If~$w_1,w_2\in\im(\finit)$, by hypothesis $w_1$ and $w_2$ have a transition in common (since they don't have internal states) hence~$w_1=w_2$.
    Let us now treat the case where~$w_1,w_2\in\im(\eta)$.
    If~$w_1$ and~$w_2$ have an internal state in common $\annot{a}{e,i}$ for some~$a\in\alphabetof\Ac$, $e\in\transitionsof{\Ac}$ and~$i\in\set{0,\ldots,m}$, which
    implies that~$w_1=w_2=\eta(e)$.
    Otherwise,~$w_1$ and~$w_2$ have a transition in common, and since they are both of length~$m+2$ it also means that they have an internal state in common and we may apply the previous case.

    This concludes the proof of the main statement of \rtheorem{topo-codi-glus}.
    We now show the second part.

    \begin{lemma}
        In $R$ there are exactly $(\card(Q)+m(m+1))$ atoms and 
        $(\card(I)+\card(F))$ occurrences of $\varepsilon$.
    \end{lemma}
    \begin{proof}
    Let~$e=(s,a,t)\in\transitionsof{\Ac}$.
    It gives rise to two subexpressions in $R_\Ac$: 
    $a^{H(e)}$ in $R^\text{right}_s$ and $a^{G(e)}$ in $R^\text{left}_t$.
    In total, $H(e)+G(e)=m+1$ atoms.
    Moreover, there are exactly $\card(Q)$ occurrences of $\sigma$
    and $(\card(I)+\card(F))$ occurrences of~$\varepsilon$.
    \end{proof}
    
    Since~$\Ac$ is trim,~$\card(Q)\leq m$ hence Lemma~\thetheorem{} yields that the size of~$R$ is in~$O(m^2)$.
    
    \begin{lemma}
        The only states in~$\gl(R)$ that have more than one outgoing edges are in
        \begin{equation}
            \im(\nu)\cup \setst*{\annot{a}{e,H(e)-1}}{e\in\transitionsof{\Ac}\text{ and } a=\lbl(e)}
        \end{equation}
        
        The only states in~$\gl(R)$ that have more than one incoming edges are in
        \begin{equation}
            \im(\nu)\cup \setst*{\annot{a}{e,H(e)}}{e\in\transitionsof{\Ac}\text{ and } a=\lbl(e)}
        \end{equation}
    \end{lemma}
    
    \begin{corollary}
        The number of transitions in~$\gl(R)$ is in~$O(m^2)$.
    \end{corollary}

    
    

\section{Application of Theorem \ref{t.topo-codi-glus} to the proof of
Theorem \ref{t.glus-does-not-help}}
\lsection{glus-does-not-help-proo}

\rtheorem{glus-does-not-help} follows from Theorems \rtheorem*{walk-memb-untr} and \rtheorem*{topo-codi-glus}, together with the next proposition. Note that the database in the proof of
\rtheorem{walk-memb-trail-trac} is simply-labelled: We say that a database~$D=(\Sigma, V, E,\src,\tgt,\lbl)$ is \emph{simply-labelled} if
$\card(\lbl(v))=1$ for every~$v\in V$.

\begin{proposition}
    Let~$\Ac$ and~$\Bc$ be two automata such that~$\Bc$ is a topological coding of~$\Ac$.
    Let~$D$ be a simply-labelled database and~$w$ a walk in~$D$.
    There exists a simply-labelled database~$D'$ and a walk~$w'$ such that \problemfont{Walk Membership} returns true on~$D,w,\Ac$ if and only if \problemfont{Walk Membership} returns true on~$D',w',\Bc$.
\end{proposition}
\begin{proof} 
We reuse notation from \rdefinition{topo-codi} for~$u_i,u_f,\lambda,\nu,\finit,$ $\eta,\ffinal$.
The database $D'=(\Sigma_\Bc, V', E',\src',\tgt',\lbl')$ is built from the database~$D=(\Sigma_\Ac, V, E,\src,\tgt,\lbl)$ as follows:
   \begin{itemize}
       \item $V'$ contains $V$
       \item each edge~$e\in E$ is replaced in $D'$ by a walk~$w_e$ with label~$\lambda(\lbl(e))$:
       $\len(\lambda(i))$ fresh edges are added to $E'$ and $\len(\lambda(i))-1$ fresh nodes are added to~$V'$ for each~$e\in E$.
       \item we add in~$D'$ one walk $w_i$ from a fresh node $S$ to $\src(w)$ and one walk $w_f$ from~$\tgt(w)$ to a fresh node $T$:
       \begin{gather*}
           w_i = \textsf{S}\rarrow{u_i}src(w)\\
           w_f = \tgt(w)\rarrow{u_f}\textsf{T}\\
       \end{gather*}
       That is $(\len(u_i)+\len(u_f))$ fresh nodes and edges.
   \end{itemize}
   The walk $w'$ is built from~$w$ as follows: we let~$e_1,e_2,\ldots,e_k$ be the edges in~$w$
   and:
   \begin{equation*}
       w' = w_i \cdot w_{e_1}\cdot w_{e_2}\cdots w_{e_k} \cdot w_f
   \end{equation*}

    \medskip

    Assume that there is~$r\in\sem[SR]{\Ac}(D)$ such that~$\projD(r)=w$.
    In the following, we denote the vertices in the run database in column to improve readability; typically $(n,s)$ is written $\bistack{n}{s}$.
    We denote~$r$ and~$w$ as follows.
    \begin{align}
        r={}&\bistack{n_0}{s_0} \rarrow{a_1} \bistack{n_1}{s_1} \rarrow{a_2}
        \cdots \rarrow{a_k} \bistack{n_k}{s_k}\\
        w ={}& (n_0,e_1,n_1,\cdots,e_k,n_k)
    \end{align}
    with~$s_0\in\initialsof\Ac$ and~$s_0\in\finalsof\Ac$.
    Hence~$\pi_\Ac$, below, is a successful computation in~$\Ac$.
    \begin{gather}
        \pi_\Ac= s_0\rarrow{a_1}s_1\rarrow{a_2}\cdots \rarrow{a_k} s_k
        \\
    \shortintertext{We denote:}
        \forall i, 0<i\leq k,\quad \delta_i= (n_{i-1},a_i,n_{i})
    \end{gather}
    Hence the corresponding computation (\rdefinition{corr-comp}), denoted by~$\pi_\Bc$ and given below, is successful in~$\Bc$ (from \rlemma{corr-run-succ}).
\begin{equation*}\newcommand{\myrule}{\rule[-2.5ex]{0pt}{6ex}}
        \overbrace{\myrule x\rarrow{u_i}\nu(s_0)}^{\finit(s_0)}
        \setlength{\dimen0}{0pt-\widthof{$\nu(s_0)$}}\hspace{\dimen0}
        \underbrace{\myrule\setlength{\dimen0}{\widthof{$\nu(s_0)$}}\hspace*{\dimen0}\rarrow{\lambda(a_1)}\nu(s_1)}_{\eta(\delta_1)\mathrlap{\hspace*{22mm}\displaystyle\cdots}}\rarrow{\lambda(a_2)}~\cdots~\rarrow{\lambda(a_{k-1})} 
        \underbrace{\myrule\rule[-1.8ex]{0pt}{0pt}\nu(s_{k-1}) \rarrow{\lambda(a_k)} \nu(s_k)}_{\eta(\delta_k)}
        \setlength{\dimen0}{0pt-\widthof{$\nu(s_k)$}}\hspace{\dimen0}\overbrace{\myrule\setlength{\dimen0}{\widthof{$\nu(s_k)$}}\hspace*{\dimen0}\rarrow{u_f} y}^{\ffinal(s_k)}
\end{equation*}
    where~$x=\src(i(s_0))$ and~$y=\tgt(f(s_k))$.
    Hence $r'$, defined below, is a run in~$D'\times\Bc$.
    \begin{equation*}
        r' = \bistack{\mathsf{S}}{x} 
        \rarrow[\finit(s_0)]{w_i}\bistack{n_0}{\nu(s_0)} 
        \rarrow[\eta(\delta_1)]{w_{e_1}} \bistack{n_1}{\nu(s_1)}
        \cdots \rarrow[\eta(\delta_k)]{w_{e_k}}
        \bistack{n_k}{\nu(s_k)}
        \rarrow[\ffinal(s_k)]{w_f}\bistack{\mathsf{T}}{y} 
    \end{equation*}
    It remains to show that~$r'$ is simple. 
    We assume that it is not simple for the sake of contradiction; let~$N,M$ be two vertices in~$r'$ such that~$N=M$.
    
    \bigskip
    
    \enumstyle{1} If~$N=\bistack{S}{x}$ then no vertex~$M$ can be equal to it since~$S$ is a fresh vertex in~$D'$: it does not occur in~$w_{i}$ or $w_f$, nor in any~$w_{e_j}$.
    
    \enumstyle{2} The three following cases are treated in the same way:\\ \enumstyle{2a} $N=\bistack{T}{y}$; \enumstyle{2b} $N$ is an internal node of $\rarrow[\finit(s_0)]{w_i}$; and \enumstyle{2c} $N$ is an internal node of $\rarrow[\ffinal(s_k)]{w_f}$.
    
    \enumstyle{3} Case where~$N=\bistack{n_i}{\nu(s_i)}$ and $M=\bistack{n_j}{\nu(s_j)}$ for some~$i,j$. 
    Since~$\nu$ is a bijection, 
    $N=M$
    implies~$\bistack{n_i}{s_i}=\bistack{n_j}{s_j}$, hence~$i=j$ since~$r$ is simple.
    
    \bigskip

    \bigskip
    
    \enumstyle{4} Case where $N$ is an internal vertex in $\rarrow[\eta(\delta_i)]{w_{e_i}}$ for some~$i$
    and~$M$ is an internal vertex in $\rarrow[\eta(\delta_j)]{w_{e_j}}$ for some~$j$.
    It implies that $\eta(\delta_i)$ and $\eta(\delta_j)$ have an internal state in common,
    hence that $\eta(\delta_i)=\eta(\delta_j)$ from \rdefinition{topo-codi-inje},
    hence that~$\nu(s_i)=\nu(s_j)$.    %
    Similarly, the internal vertices in~$w_{e_i}$ and~$w_{e_j}$ were created fresh, hence $w_{e_i}=w_{e_j}$. It follows that~$n_i=n_j$.
    
    Finally, we have $\bistack{n_i}{\nu(s_i)}=\bistack{n_j}{\nu(s_j)}$ and we apply case \enumstyle{3}.
    
    \bigskip
    
    \enumstyle{5} The last case is where, for some~$i,j$, $N$ is an internal vertex in $\rarrow[\eta(\delta_i)]{w_{e_i}}$ and $M=\bistack{n_j}{\nu(s_j)}$.
    It would implies that an internal node in~$w_{e_i}$, which is a fresh node in~$D'$, is equal to~$n_j$, which was already in~$D$, a contradiction.
    
    \bigskip

    It remains to show the converse: the existence of a simple run~$r$ in $D'\times\Bc$ implies the existence of a simple run~$r'$ in~$D\times\Ac$.
    We use \rlemma{inv-run-succ} to build a successful computation in~$\Ac$ from the successful computation in~$\Bc$ underlying~$r$, and then we build a run~$r'\in D\times\Ac$.
    In that direction, showing that~$r'$ is simple is directly implied by the fact that~$r$ is simple.
\end{proof}

\section{About \protect\rproposition{bindingtrail-simplerun-equivalence}}
\lsection{proof-prop}

The technique developed earlier in appendix~\theappendix{} allows to prove \rproposition{bindingtrail-simplerun-equivalence}, recalled below.

\begin{falsestatement}{\rproposition{bindingtrail-simplerun-equivalence}}
    \bindingtrailsimplerunequivalencesttmt
\end{falsestatement}

Indeed, binding trail semantics is closely linked to the Glushkov automaton, and we show next how to use  \rtheorem{glus-does-not-help} to show one of reductions required 
for \rproposition{bindingtrail-simplerun-equivalence}.
Other reductions require similar classical graph techniques.

\begin{proposition}
    There exists a fixed expression~$R$ such that \problemfont{Walk Membership} is NP-hard for~binding-trail semantics.
\end{proposition}
\begin{proof}
    Let~$R$ be the expression and~$D=(\Sigma, V, E,\src,\tgt,\lbl)$ the database given by \rtheorem{glus-does-not-help}.
    Let~$D'$ be the database constructed from~$D$
    by splitting vertices in order for every vertex in~$D'$ to have at most one incoming transition. More precisely~$D'=(\Sigma, V'\cup V, E'\cup E,\src',\tgt',\lbl')$ where
    \begin{itemize}
        \item $V' = \setst{(e,v)\in E\times V}{\tgt(e)=v}$;
        \item $E' = \setst{((e,v),e')\in V'\times E}{\src(e')=v}$;
        \item $\src'(((e,v),e')) = (e,v)$; \quad $\src'(e)=\src(e)$;
        \item $\tgt'(((e,v),e')) = (e',\tgt(e'))$; \quad $\tgt'(e)=(e,\tgt(e))$;
        \item $\lbl'(((e,v),e')) = \lbl(e')$; \quad $\src'(e)=\src(e)$.
    \end{itemize}
    
    The simple runs in $D\times\gl(R)$ are in bijection with the binding trails in~$D$ matching~$R$ that start with an edge in~$E$.
    Indeed, the simple-run $r=((v_0,s_0),(e_1,t_1),(v_1,s_1),\ldots,(e_n,t_n),(v_n,s_n))$ in~$D\times\gl(R)$ is associated with:
    $(f_1,\alpha_1)\cdots(f_n,\alpha_n)$, where 
    \begin{itemize}
        \item $f_1=e_1$, $f_i=((e_{i-1})$ for every~$i$, $1<i\leq n$; and
        \item $\alpha_n$ is the letter labelling transition~$t_n$ for every~$i$, $0<i\leq n$.
    \end{itemize}
    One may verify that tfae
    \begin{itemize}
        \item $(f_i,\alpha_i)=(f_j,\alpha_j)$
        \item $(v_i,s_i)=(v_j,s_j)$
    \end{itemize}
    Note also that $(v_0,s_0)=(v_i,s_i)$ implies~$i=0$:
    the state~$s_0$ is necessarily the special initial state from the Glushkov Construction and thus has no incoming transition.
\end{proof}

%% file: appendix/proof_cypherlike.tex
\newappendix{Proof of \rtheorem{cypherlike-ptime-path-memb}}
\label{a:cypherlike}

The purpose of Appendix~\theappendix{} is to show \rtheorem{cypherlike-ptime-path-memb}, restated below.

\cypherlikethm*

We only show the statement for Glushkov automaton; the proof is similar for binding-trail semantics.

\medskip

First, \rlemma{cyph-rest} states that as soon as we don't allow concatenation under star, the expression may be simplified syntactically.

\begin{lemma}\llemma{cyph-rest}
    Let~$R$ be an expression with no concatenation under star.
    Then,
    \begin{itemize}
        \item deleting all stars that are nested inside another star,
        \item deleting every occurrence of~$\varepsilon$ that appears inside a star
    \end{itemize}
    yields an expression~$R'$ such that~$\gl(R)=\gl(R')$.
\end{lemma}

\rlemma{cyph-rest} describes a much simplified version of the algorithm to put an expression in \emph{star-normal form} \cite{BrugemannKlein93,Sakarovitch2021}.

\begin{proposition}
    \problemfont{Walk Membership} under simple-run semantics is in P-time if the input is~$\Ac=\gl(R)$ where~$R$ is an expression with no concatenation under star.
\end{proposition}
\begin{proof}
    Let~$R$ be an expression with no concatenation under star.
    From \rlemma{cyph-rest}, we may assume that~$R$ has no star nor occurrence of $\varepsilon$ inside a star.
    Let~$w=(n_0,e_1,n_1,\ldots,e_m,n_m)$ be a walk in the 
    database~$D = (\Sigma, V, E,\src,\tgt,\lbl)$.
    
    The general strategy is to compute inductively the
    set~$S_X$ given below, for each subexpression~$X$ of~$R$.
    \begin{equation*}
        S_X = \setst*{(i,j)}{%
            \begin{array}{@{}l@{}}
            0\leq i \leq j \leq m\\
            (n_i,e_{i+1},\ldots,n_j) \in\sem[SR]{\gl(X)}(D)
            \end{array}
        }
    \end{equation*}
    
    First, $S_{X\cdot Y}$ and $S_{X+Y}$ are easy to compute in polynomial time
    from~$S_X$ and~$S_Y$.
    Second, $S_\varepsilon=\setst{(i,i)}{0\leq i \leq m}$ and, for each~$a\in\Sigma$ $S_a=\setst{(i,i+1)}{0\leq i <m\text{ and }a\in\lbl(e_{i+1})}$ is built in linear time.
    The remainder of the proof is about the last case, that is where~$X=(a_1 + \cdots + a_n)^*$, for some atoms~$a_1,\ldots, a_n$.
    Note that it is possible that~$a_i=a_j$ for some~$i\neq j$.
    
    \medskip
    
    We build~$S_X$ by testing whether~$(\ell,k) \in S_X$ for each
    $\ell,k$ such that~$0\leq\ell\leq k \leq m$.
    We now describe a polynomial time algorithm to test whether~$(\ell,k) \in S_X$.
    For each vertex~$v$ in~$w$, we let~$I_v$ denote the set~$I_v=\setst[\big]{i\in\set{\ell,\ldots,k-1}}{\tgt(e_i)=v}$.
    Consider the following undirected graph~$H_v=(V,U)$
    \begin{itemize}
        \item $V$ contains~$\card(I_v)+n$ vertices, one vertex~$P_i$ for each position~$i$ in~$I_v$, plus one vertex~$A_j$ for each atom~$a_j$:
        \begin{equation*}
            V=\setst{P_i}{i\in I_v}\cup\setst{A_j}{0< j \leq n}
        \end{equation*}
        \item $U$ contains an edge between $P_i$ and $A_j$ if and only if $a_j$ is a label of $e_i$:
        \begin{equation*}
            U=\setst*{(P_i,A_j)}{\begin{array}{@{}l@{}}
            i\in I_v\\0<j\leq n\\
            a_j\in\lbl(e_i)\end{array}}
        \end{equation*}
    \end{itemize}
    Note that~$H_v$ is a bipartite graph.
    We may then use a classical algorithm to compute the maximal matching~$M_v$ of~$H_v$ in polynomial time \cite[Section~26.3]{CormenEtAl2009}\cite{HopcroftKarp1973}.
    
    Then, one may use the different $M_v$'s to test whether $(\ell,k)\in S_X$, as stated below.
    
    \begin{claim} The following are equivalent.
    \begin{subthm}
        \item For each~$i$, $\ell<i\leq k$, $\card(M_{n_i})=\card(I_{n_i})$;
        \item $(\ell,k)\in S_X$.
    \end{subthm}
    \end{claim}
    \begin{proofofclaim}
        $\enumstyle{a}\Rightarrow\enumstyle{b}$.
        Let us consider the automaton~$\gl(X)$. 
        It has~$(n+1)$ states:
        one initial state~$q_0$, plus one state~$q_j$ for each atom~$a_j$, $1\leq j\leq n$.  
        All states are final.
        It has~$(n+1)n$ transitions: $(q_i, a_j, q_j)$ for each~$i,j$ such that~$0\leq i\leq n$ and $0<j\leq n$.
        
        Let us construct a run~$r$ such that~$\projD(r)=(n_\ell,e_{\ell+1},\ldots,n_k)$.
        It is defined by
        \begin{equation*}
            r=
            \bistack{n_\ell}{s_\ell} 
            \rarrow[b_{\ell+1}]{e_{\ell+1}}
            \bistack{n_{\ell+1}}{s_{\ell+1}}
            \rarrow[b_{\ell+2}]{e_{\ell+2}}
            \cdots
            \rarrow[b_k]{e_k}
            \bistack{n_{k}}{s_{k}}
        \end{equation*}
        where~$s_\ell=q_0$ and for every~$i$, $\ell<i\leq k$, we set $s_i=q_j$,
        and~$b_i=a_j$, where~$j$ is the index of the vertex~$A_j$ matched to vertex~$P_i$ in~$M_{n_i}$.
        By construction~$s_\ell\rarrow{b_\ell+1}\cdots \rarrow{b_k}s_{k}$ is an (accepting) computation in~$\gl(X)$.
        For the sake of contradiction, assume that~$r$ is not simple.
        There exists~$i,i',j$ such that~$n_{i}=n_{i'}$ and~$s_{i}=s_{i'}=q_j$
        hence by definition, $A_j$ is matched to both~$P_i$ and~$P_i'$ in the matching~$M_{n_i}$,
        a contradiction.
        
        The proof of $\enumstyle{b}\Rightarrow\enumstyle{a}$ is similar: the construction of the run from the matchings is actually bijective. (The matchings built from the run are necessarily maximal since they use all $P_i$'s.)
    \end{proofofclaim}
    
    In order to compute~$S_X$, we use the algorithm above for each pair~$(k,\ell)$, $0\leq k < \ell \leq m$; which results in a polynomial time algorithm overall.    
    The number of subexpressions of~$R$ are in polynomial number so computing
    all~$S_X$'s may be done in polynomial time, and then one simply has to check that~$S_R$ contains~$(0,m)$.
\end{proof}

%% file: appendix/trailrun.tex
\newappendix{Trail-Run Semantics}

Trail semantics is quite spread through the language Cypher, while simple walk semantics seems to be less popular.
Hence, one could ask why we chose to define a semantics based on the simple walks in the run database.
We could indeed define a semantics based on trails and we discuss in this section why it seems to yield a worse semantics.

\begin{definition}[Trail-run semantics]\ldefinition{trail-run semantics}
    The \emph{trail-run semantics} of an automaton $\Ac$, denoted by~$\sem[SR]{\Ac}$, is the mapping that associates, to each database $D$, the following bag of answers:
    \begin{align*}
        &\sem[SR]{\Ac}(D) = \projD\compose\trail\compose\match{\Ac}{D}~.
    \end{align*}
\end{definition}

Note that $\trail$ is a coarser filter than~$\simple$, hence the number of results will always be greater when using trail-run semantics rather that simple-run semantics (see \rfigure{sema-incl}).
It could seem better to obtain more result, but recall that we define here results that are deemed \emph{redundant} and it seems to us that the results filtered out by simple-run semantics, but kept in trail-run semantics usually seem to be uninteresting and to grow quite fast in number in pathological cases.

\begin{figure}[ht]\centering
\subfloat[One loop database]{~~~~\input{figure/one_loop_db.tex}~~~~\lfigure{one-loop-db}}%
\hfil\hfil%
\subfloat[One state automaton with several loops]{~~~~\input{figure/one_loop_aut.tex}~~~~\lfigure{one-loop-aut}}%
\hfil\hfil%
\subfloat[Standard]%
{~~~~\input{figure/one_loop_std_aut.tex}~~~~\lfigure{one-loop-std-aut}}%
\caption{Pathological case for trail-run semantics}
\lfigure{trail-run-big}
\end{figure}

For instance, consider the database and automaton given in \rfigure{trail-run-big}. 
With simple-run semantics, only the empty walk will be returned.
With trail-run semantics, one path consisting of $i$-times the edge $e$ will be returned for every~$i$, $0\leq i\leq n$: the length-$i$ path is returned with multiplicity~$(C_i^n\times i!)$, and the total number of returned walks is in~$O(n(n!))$.

Another example is the Glushkov automaton associated with~$(a_1+\cdots+a_n)^*$.
It consists of one initial state plus a clique of size~$n$. 
Hence the total number of returned walks is in~$O(n(n!))$ for simple-run semantics and~$O(n^3(n^2!))$.)

More generally, trail-run semantics enforces that every specific transition of the automaton is used once (per edge in the graph).  However, in most automata, transitions have no particular meaning.
Indeed, if a computation uses twice the same transition, we may write it as:
\begin{equation*}
    s_0\rarrow{v1} s_1 \rarrow{a} s_2 \rarrow{v_2} s_3\rarrow{b} s_4\rarrow{v3} s_5\quad\quad\text{with}\quad a,b\in\Sigma \quad v_1,v_2,v_2\in\Sigma^*;
\end{equation*}
and such that$s_1=s_3$, $s_2=s_3$, and $a=b$.
In essence, the computation has to use the same state twice in a row.

For instance, the transitions in Glushkov automata have no particular meaning,
and trail-run semantics make worse some issues previously mentioned, such as the moderate left-to-right bias (cf. \rsection{todo}).

Consider the following expression, parametrized by two positive integers~$i$ and~$j$.
\begin{equation*}
    E_{i,j} = ((a_1+\cdots+a_i) b(c_1+\cdots+c_j))^*
\end{equation*}
It is clearly symmetric but the number of times a letter~$c_k$ may match the same edge in the database is always 1 while the number of times a letter~$a_k$ may match the same edge is equal to~$j$.
This example highlights the fact that the number of transitions created by the Glushkov construction from one particular occurrence of a letter does not bear any
particular meaning.  
It is highly dependent on on the remainder of the expression.

If we allow forward and backward edges, the two following would not be equivalent:
\begin{align*}
    \overleft{E_{i,j}} &{}= \left((\overleft{a_1}+\cdots+\overleft{a_i}) \overleft{b}(\overleft{c_1}+\cdots+\overleft{c_j})\right)^* \\
    \overright{E_{i,j}} &{}= \left((\overright{c_1}+\cdots+\overright{c_j}) \overright{b}(\overright{a_1}+\cdots+\overright{a_i})\right)^* \\
\end{align*}

%% file: figure/one_loop_db.tex
\begin{tikzpicture}
\node[node] (n1) {1};

\path[north loop] (n1) to node[above] {$e:\{a_1,\cdots,a_n\}$} (n1);
\end{tikzpicture}

%% file: figure/one_loop_aut.tex
\begin{tikzpicture}

\node[node] (n1) {1};
\initialfinal[180]{n1}

\path[north east loop] (n1) to node[right] (a1) {$a_1$} (n1);
\path[south east loop] (n1) to node[right] (a2) {$a_n$} (n1);
\path (a1) to node[midway,anchor=center] {$\cdot$} (a2);
\path (a1) to node[pos=.3,anchor=center] {$\cdot$} (a2);
\path (a1) to node[pos=.7,anchor=center] {$\cdot$} (a2);

\end{tikzpicture}

%% file: figure/one_loop_std_aut.tex
\begin{tikzpicture}

\node[node] (n1) {1};
\initial[90]{n1}
\final[-90]{n1}

\path (n1) ++(0:\nodedist) node[node] (n2) {2} 
;

\final[-90]{n2}

\path[edge] (n1) to[bend left]  node[above] (b1) {$a_1$} (n2);
\path[edge] (n1) to[bend right] node[below] (b2) {$a_n$} (n2);
\path (b1) to node[midway,anchor=center] {$\cdot$} (b2);
\path (b1) to node[pos=.3,anchor=center] {$\cdot$} (b2);
\path (b1) to node[pos=.7,anchor=center] {$\cdot$} (b2);

\path[north east loop] (n2) to node[right] (a1) {$a_1$} (n2);
\path[south east loop] (n2) to node[right] (a2) {$a_n$} (n2);
\path (a1) to node[midway,anchor=center] {$\cdot$} (a2);
\path (a1) to node[pos=.3,anchor=center] {$\cdot$} (a2);
\path (a1) to node[pos=.7,anchor=center] {$\cdot$} (a2);

\end{tikzpicture}

%% file: main.bbl
\begin{thebibliography}{10}

\bibitem{AnglesEtAl2018}
Renzo Angles, Marcelo Arenas, Pablo Barcel{\'o}, Peter~A. Boncz, George H.~L.
  Fletcher, Claudio Gutierrez, Tobias Lindaaker, Marcus Paradies, Stefan
  Plantikow, Juan~F. Sequeda, Oskar van Rest, and Hannes Voigt.
\newblock {G-CORE:} {A} core for future graph query languages.
\newblock In {\em {SIGMOD}}, pages 1421--1432. {ACM}, 2018.

\bibitem{AnglesEtAl2017}
Renzo Angles, Marcelo Arenas, Pablo Barcel{\'{o}}, Aidan Hogan, Juan~L.
  Reutter, and Domagoj Vrgoc.
\newblock Foundations of modern query languages for graph databases.
\newblock {\em {ACM} Comput. Surv.}, 50(5), 2017.

\bibitem{BaganBonifatiGroz2020}
Guillaume Bagan, Angela Bonifati, and Beno{\^{\i}}t Groz.
\newblock A trichotomy for regular simple path queries on graphs.
\newblock {\em J. Comput. Syst. Sci.}, 108:29--48, 2020.

\bibitem{BonifatiMartensTimm2020}
Angela Bonifati, Wim Martens, and Thomas Timm.
\newblock An analytical study of large {SPARQL} query logs.
\newblock {\em {VLDB} J.}, 29(2-3):655--679, 2020.

\bibitem{BrugemannKlein93}
Anna Br\"{u}gemann-Klein.
\newblock Regular expressions into finite automata.
\newblock {\em Theoretical Computer Science}, 120:197--213, 1993.

\bibitem{CaronZiadi2000}
Pascal Caron and Djelloul Ziadi.
\newblock Characterization of glushkov automata.
\newblock {\em Theor. Comput. Sci.}, 233(1-2):75--90, 2000.

\bibitem{SPARQL1.1PP}
World Wide~Web Consortium.
\newblock Section 9: Property paths.
\newblock In {\em SPARQL 1.1 Query Language}. 2013.
\newblock \url{https://www.w3.org/TR/sparql11-query/\#propertypaths}.

\bibitem{CormenEtAl2009}
T.H. Cormen, C.E. Leiserson, R.L. Rivest, and C.~Stein.
\newblock {\em Introduction to Algorithms, third edition}.
\newblock MIT Press, 2009.

\bibitem{CruzMendelzonWood1987}
Isabel~F. Cruz, Alberto~O. Mendelzon, and Peter~T. Wood.
\newblock A graphical query language supporting recursion.
\newblock In Umeshwar Dayal and Irving~L. Traiger, editors, {\em Proceedings of
  the Association for Computing Machinery Special Interest Group on Management
  of Data 1987 Annual Conference, San Francisco, CA, USA, May 27-29, 1987},
  pages 323--330. {ACM} Press, 1987.
\newblock \href {https://doi.org/10.1145/38713.38749}
  {\path{doi:10.1145/38713.38749}}.

\bibitem{DeutschEtAl2022}
Alin Deutsch, Nadime Francis, Alastair Green, Keith Hare, Bei Li, Leonid
  Libkin, Tobias Lindaaker, Victor Marsault, Wim Martens, Jan Michels, Filip
  Murlak, Stefan Plantikow, Petra Selmer, Hannes Voigt, Oskar van Rest, Domagoj
  Vrgo\v{c}, Mingxi Wu, and Fred Zemke.
\newblock Graph pattern matching in {GQL} and {SQL}/{PGQ}, 2022.

\bibitem{DeutschEtAl2019}
Alin Deutsch, Yu~Xu, Mingxi Wu, and Victor~E. Lee.
\newblock Tigergraph: {A} native {MPP} graph database, 2019.
\newblock arXiv preprint.
\newblock URL: \url{http://arxiv.org/abs/1901.08248}.

\bibitem{DBLP:books/daglib/0030488}
Reinhard Diestel.
\newblock {\em Graph Theory, 4th Edition}, volume 173 of {\em Graduate texts in
  mathematics}.
\newblock Springer, 2012.

\bibitem{GQL-ISO}
International~Organization for Standardization.
\newblock Gql.
\newblock Standard under development ISO/IEC CD 39075, March 2023 (expected).
\newblock URL: \url{https://www.iso.org/standard/76120.html}.

\bibitem{FrancisEtAl2018}
Nadime Francis, Alastair Green, Paolo Guagliardo, Leonid Libkin, Tobias
  Lindaaker, Victor Marsault, Stefan Plantikow, Mats Rydberg, Petra Selmer, and
  Andr\'es Taylor.
\newblock Cypher: An evolving query language for property graphs.
\newblock In {\em SIGMOD'18}. ACM, 2018.

\bibitem{HopcroftKarp1973}
John~E. Hopcroft and Richard~M. Karp.
\newblock An $n^{5/2}$ algorithm for maximum matchings in bipartite graphs.
\newblock {\em SIAM Journal on Computing}, 2(4):225--231, 1973.

\bibitem{MartensNiewerthTrautner2020}
Wim Martens, Matthias Niewerth, and Tina Trautner.
\newblock A trichotomy for regular trail queries.
\newblock In Christophe Paul and Markus Bl{\"{a}}ser, editors, {\em STACS'20},
  volume 154 of {\em LIPIcs}, pages 7:1--7:16. Schloss Dagstuhl -
  Leibniz-Zentrum f{\"{u}}r Informatik, 2020.
\newblock \href {https://doi.org/10.4230/LIPIcs.STACS.2020.7}
  {\path{doi:10.4230/LIPIcs.STACS.2020.7}}.

\bibitem{MendelzonWood1995}
Alberto~O. Mendelzon and Peter~T. Wood.
\newblock Finding regular simple paths in graph databases.
\newblock {\em {SIAM} J. Comput.}, 24(6):1235--1258, 1995.

\bibitem{RobinsonWebberEifrem2015}
Ian Robinson, Jim Webber, and Emil Eifrem.
\newblock {\em Graph databases, second edition}.
\newblock O'Reilly, 2015.

\bibitem{Sakarovitch2021}
Jacques Sakarovitch.
\newblock Automata and rational expressions, 2021.
\newblock Chapter 2 of \cite{Pin2021}.

\bibitem{TigerGraph3.1}
TigerGraph Team.
\newblock {TigerGraph Documentation -- version 3.1}, 2021.
\newblock URL: \url{https://docs.tigergraph.com/}.

\bibitem{Valiant1979}
Leslie~G. Valiant.
\newblock The complexity of enumeration and reliability problems.
\newblock {\em SIAM J. Comput.}, 8(3), 1979.

\bibitem{Yen71}
Jin~Y. Yen.
\newblock Finding the k shortest loopless paths in a network.
\newblock {\em Management Science}, 17(11):712--716, 1971.
\newblock URL: \url{http://www.jstor.org/stable/2629312}.

\end{thebibliography}
